\RequirePackage{fix-cm}
\documentclass[twocolumn, epjc3, comma, sort&compress, natbib]{svjour3}
\bibpunct{[}{]}{,}{n}{}{,} 

\journalname{Eur. Phys. J. A}

\usepackage{latexsym}
\usepackage{amsmath}
\usepackage{amssymb}
\usepackage{amsfonts}
\usepackage{CJKutf8}
\usepackage{enumitem}

\usepackage[mathscr,scaled=1.15]{urwchancal}
\DeclareFontFamily{OT1}{pzc}{}
\DeclareFontShape{OT1}{pzc}{m}{it}%
{<-> s * [1.15] pzcmi7t}{}
\DeclareMathAlphabet{\mathpzc}{OT1}{pzc}{m}{it}

\usepackage{color}

\usepackage{supertabular}
\usepackage{placeins}
\usepackage{epsfig}
\usepackage{graphicx}

\definecolor{purple}{rgb}{0.5,0,0.5}
\definecolor{blue}{rgb}{0.0,0,0.9}
\definecolor{prdblue}{rgb}{0.133,0.118,0.498}
\usepackage[colorlinks=true, pdfstartview=FitV, linkcolor=prdblue, citecolor= prdblue, urlcolor=prdblue]{hyperref}

\begin{document}
\begin{CJK*}{UTF8}{gbsn}

\title{$\,$\\[-6ex]\hspace*{\fill}{\normalsize{\sf\emph{Preprint no}.\
NJU-INP 104/25}}\\[1ex]
Quark + Diquark Description of Nucleon Elastic Electromagnetic Form Factors}

\author{
    Peng Cheng (程鹏)\thanksref{AHNU}%
    $\,^{\href{https://orcid.org/0000-0002-6410-9465}{\textcolor[rgb]{0.00,1.00,0.00}{\sf ID}}}$
\and
Zhao Qian Yao (姚照千)\thanksref{UHe,UPO}%
       $\,^{\href{https://orcid.org/0000-0002-9621-6994}{\textcolor[rgb]{0.00,1.00,0.00}{\sf ID}}}$
\and
    Daniele Binosi\thanksref{ECT}%
    $\,^{\href{https://orcid.org/0000-0003-1742-4689}{\textcolor[rgb]{0.00,1.00,0.00}{\sf ID}}}$
%
\and \\Ya Lu (陆亚)\thanksref{NJT}%
       $\,^{\href{https://orcid.org/0000-0002-0262-1287}{\textcolor[rgb]{0.00,1.00,0.00}{\sf ID}}}$
%
\and
    Craig D.\ Roberts\thanksref{NJU,INP}%
       $\,^{\href{https://orcid.org/0000-0002-2937-1361}{\textcolor[rgb]{0.00,1.00,0.00}{\sf ID}}}$
}

\institute{Department of Physics, \href{https://ror.org/05fsfvw79}{Anhui Normal University}, Wuhu, Anhui 24100, China\label{AHNU}
\and
Dpto.~Ciencias Integradas, Centro de Estudios Avanzados en Fis., Mat. y Comp., \\
\hspace*{0.5em}Fac.~Ciencias Experimentales, \href{https://ror.org/03a1kt624}{Universidad de Huelva}, E-21071 Huelva, Spain
\label{UHe}
\and
Dpto. Sistemas F\'isicos, Qu\'imicos y Naturales, Univ.\ \href{https://ror.org/02z749649}{Pablo de Olavide}, E-41013 Sevilla, Spain
\label{UPO}
\and
European Centre for Theoretical Studies in Nuclear Physics
            and Related Areas  (\href{https://ror.org/01gzye136}{ECT*})\\ \hspace*{0.5em}Villa Tambosi, Strada delle Tabarelle 286, I-38123 Villazzano (TN), Italy\label{ECT}
\and Department of Physics, \href{https://ror.org/03sd35x91}{Nanjing Tech University}, Nanjing 211816, China \label{NJT}
\and
School of Physics, \href{https://ror.org/01rxvg760}{Nanjing University}, Nanjing, Jiangsu 210093, China \label{NJU}
\and
Institute for Nonperturbative Physics, \href{https://ror.org/01rxvg760}{Nanjing University}, Nanjing, Jiangsu 210093, China
\label{INP}
%
%
           %
\\[1ex]
Email:
\href{mailto:pcheng@mail.ahnu.edu.cn}{pcheng@mail.ahnu.edu.cn} (PC),
\href{mailto:zhaoqian.yao@dci.uhu.es}{zhaoqian.yao@dci.uhu.es} (ZQY),
\href{mailto:binosi@ectstar.eu}{binosi@ectstar.eu} (DB),\\
\hspace*{2.85em}
\href{mailto:luya@njtech.edu.cn}{luya@njtech.edu.cn} (YL),
\href{mailto:cdroberts@nju.edu.cn}{cdroberts@nju.edu.cn} (CDR)
            }

\date{2025 July 17}

\maketitle

\end{CJK*}

\begin{abstract}
Working with a Poincar\'e-covariant quark + diquark, $q(qq)$, Faddeev equation approach to nucleon structure, a refined symmetry preserving current for electron + nucleon elastic scattering is developed.  The parameters in the interaction current are chosen to ensure that the $q(qq)$ picture reproduces selected results from  contemporary $3$-body analyses of nucleon elastic electromagnetic form factors.  Although the subset of fitted results is small, the $q(qq)$ picture reproduces almost all the $3$-body predictions and often results in better agreement with available data.  Notably, the $q(qq)$ framework predicts a zero in $G_E^p/G_M^p$, the absence of such a zero in $G_E^n/G_M^n$, and a zero in the proton's $d$-quark Dirac form factor.
Derived $q(qq)$ results for proton flavour-separated light-front-transverse number and anomalous magnetisation densities are also discussed.  With the $q(qq)$ framework thus newly benchmarked, one may proceed to comparisons with a broader array of $3$-body results.  This may enable new steps to be made toward answering an important question, \emph{viz}.\ is the quark + fully-interacting diquark picture of baryon structure only a useful phenomenology or does it come close to expressing robust features of baryon structure?
\end{abstract}

\section{Introduction}
Pictures of the nucleon as a quark + diquark system have long been popular \cite{Anselmino:1992vg}, but the meanings of ``quark'' and ``diquark'' typically depend on the practitioner.
In the earliest such models, the quark and diquark were elementary constituent-like degrees of freedom.
Despite physics shortcomings, the simplicity of these approaches is a merit; so, such treatments are still in use today \cite{Barabanov:2020jvn}.

Our perspective is different.  We begin with the analyses in Refs.\,\cite{Cahill:1988dx, Reinhardt:1989rw, Efimov:1990uz}, which employed continuum Schwin\-ger function methods (CSMs) and exploited the pairing capacity of fermions in order to simplify the three valence-quark bound state problem, deriving therewith a Faddeev equation expressed in terms of dressed quarks and fully-interacting diquark correlations built therefrom.  First, rudimentary numerical calculations were described in Ref.\,\cite{Burden:1988dt}.  Since then, the approach has evol\-ved into a sophisticated tool that has been used to predict many baryon observables -- see, \emph{e.g}., Refs.\,\cite{Eichmann:2016yit, Burkert:2017djo, Brodsky:2020vco, Roberts:2020hiw, Chen:2018nsg, Lu:2019bjs, Cui:2020rmu, Cui:2021gzg, Chen:2021guo, Chang:2022jri, Yin:2023kom, Chen:2023zhh, Liu:2023reo, Carman:2023zke, Achenbach:2025kfx}; and contribute to the discussion of systems that may have four- or five valence-quarks \cite{Eichmann:2025tzm, Eichmann:2025gyz}.

Direct \emph{ab initio} solutions of the three valence-quark bound state problem were first described in Ref.\,\cite{Eichmann:2009qa}.  With steady growth in computational resources and algorithm improvements, quite a few more recent studies have been completed.   Recently, for instance, a unified set of predictions was delivered for nucleon electromagnetic and gravitational form factors \cite{Yao:2024uej, Yao:2024ixu}.  Such calculations nevertheless remain challenging and time consuming; consequently, \emph{inter alia}, the direct three valence body approach has not yet been widely applied to nucleon-to-resonance transition form factors \cite{Burkert:2017djo, Carman:2023zke, Achenbach:2025kfx}.

Notwithstanding that, a new hybrid path now pre\-sents itself.
Namely, with the three-body approach tes\-ted in applications to nucleon form factors on a large domain of momentum transfer squared, $Q^2/{\rm GeV}^2 \in [0,12]$, it becomes possible to update the quark + fully-interacting diquark approach, used efficaciously in \linebreak Refs.\,\cite{Chen:2018nsg, Lu:2019bjs, Cui:2020rmu, Cui:2021gzg, Chen:2021guo, Chen:2023zhh}, so that it reproduces the \emph{ab initio} results, insofar as that is possible.
Supposing the match is good, then the rejuvenated quark + diquark approach can, \emph{e.g}., be used in future to review and extend existing studies of resonance electroproduction \cite{Carman:2023zke, Achenbach:2025kfx}.
In addition, by establishing a common baseline for the three-body and quark + diquark approaches, it will become possible to compare their predictions and thereby identify and highlight empirical signals for the presence of diquark correlations within nucleons.

This analysis is arranged as follows.
Section~\ref{SecFaddeev} recapitulates the quark + diquark Faddeev equation used as the basis for the form factor calculations in Refs.\,\cite{Cui:2020rmu}.  We preserve that formulation therein, which involves both isoscalar-scalar and isovector-vector diquark correlations.
However, we make significant changes to the electromagnetic current that delivers a symmetry preserving treatment of electron + nucleon ($ e N$) scattering.  They are detailed in Sect.\,\ref{SecCurrent}.
The new current involves eight parameters: the radii of the scalar and axialvector diquarks plus six parameters that characterise the unamputated photon + quark vertex.  As explained in Sect.\,\ref{Method}, the parameters are determined via a least-squares fit to three-body results \cite{Yao:2024uej} for the proton elastic electric form factor, $G_E^p(x)$, on $x\in (0,4.5)$ and the ratio $\mu_p G_E^p(x)/G_M^p(x)$ on $x\in (2,4.5)$, where $G_M^p(x)/\mu_p$ is the $x=0$ unit-normalised proton magnetic form factor and $x=Q^2/m_N^2$, with $m_N$ the nucleon mass.
Section~\ref{SecResults} presents and discusses results for all nucleon elastic electromagnetic form factors, including the electric-to-magnetic ratios for the proton and neutron \cite{Jones:1999rz, Gayou:2001qd, Punjabi:2005wq, Puckett:2010ac, Puckett:2017flj, Riordan:2010id, Madey:2003av}.
Flavour separations of these form factors are discussed in Sect.\,\ref{SecFlavour}, which also includes analyses of light-front transverse densities derived therefrom.
Section~\ref{epilogue} presents a summary and perspective.

\begin{figure}[t]
\centerline{%
\includegraphics[clip, width=0.48\textwidth]{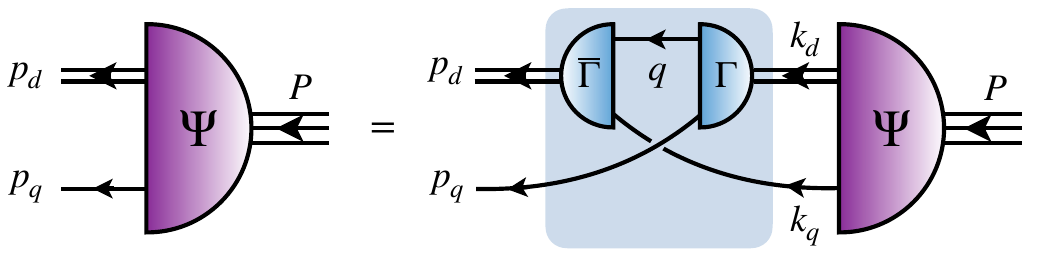}}
\caption{\label{FigFaddeev}
Integral equation for $\Psi$, the Poincar\'e-covariant matrix-valued function (Faddeev amplitude) for a nucleon with total momentum $P=p_q+p_d=k_q+k_d$ comprised of three valence quarks, two of which are contained in a fully-interacting nonpointlike diquark correlation.
$\Psi$ describes the relative momentum correlation between the dressed-quarks and -diquarks.
Legend. \emph{Shaded box} -- Faddeev kernel;
\emph{single line} -- dressed-quark propagator, Sect.\,\ref{subappendixqprop};
$\Gamma$ -- diquark correlation amplitude, Sect.\,\ref{subappendixdqbsa},
and \emph{double line} -- diquark propagator, Sect.\,\ref{subappendixdqprop}.
Only isoscalar--scalar, $[q_1q_2]$, and isovector--axialvector, $\{q_1q_2\}$, diquarks play a role in nucleons \cite{Barabanov:2020jvn}.}
\end{figure}

\section{Quark + Diquark Faddeev Equation}
\label{SecFaddeev}
In this revision and improvement of what is sometimes called the QCD-kindred approach, we do not make any changes to the Faddeev equation kernel that was used previously to calculate the nucleon's Poin\-car\'e covariant wave function; namely, it remains unchanged from that detailed in Ref.\,\cite[Sect.\,2]{Alkofer:2004yf} and widely employed subsequen\-tly; see, \emph{e.g}., Refs.\,\cite{Chen:2018nsg, Lu:2019bjs, Cui:2020rmu, Cui:2021gzg, Chen:2021guo, Chen:2023zhh}.  Herein, therefore, we present only a brief recapitulation.
(Isospin symmetry is assumed throughout.)

\subsection{Dressed quark propagator}
\label{subappendixqprop}
A principal piece of the Fig.\,\ref{FigFaddeev} Faddeev kernel is the dressed-quark propagator:
{\allowdisplaybreaks
\begin{subequations}
\begin{align}
\label{Spsigma}
S(p) & =  -i \gamma\cdot p\, \sigma_V(p^2) + \sigma_S(p^2) \\
\label{SpAB}
& = 1/[i\gamma\cdot p\, A(p^2) + B(p^2)]\,.
\end{align}
\end{subequations}
Regarding light-quarks, the wave function renormalisation and dressed-quark mass:
\begin{equation}
\label{ZMdef}
Z(p^2)=1/A(p^2)\,,\;M(p^2)=B(p^2)/A(p^2)\,,
\end{equation}
respectively, receive material momentum-dependent corrections at infrared momenta \cite{Binosi:2016wcx}: $Z(p^2)$ is suppressed and $M(p^2)$ enhanced.  These features are an expression of emergent hadron mass (EHM) \cite{Roberts:2021nhw, Ding:2022ows, Binosi:2022djx, Ferreira:2023fva, Raya:2024ejx}.

The following parametrisation of $S(p)$, which exhibits the characteristics described above, has long been used in hadron studies:
{\allowdisplaybreaks
\begin{subequations}
\label{EqSSSV}
\begin{align}
\bar\sigma_S(x) & =  2\,\bar m \,{\cal F}(2 (x+\bar m^2)) \nonumber \\
& \quad + {\cal F}(b_1 x) \,{\cal F}(b_3 x) \,
\left[b_0 + b_2 {\cal F}(\epsilon x)\right]\,,\label{ssm} \\
\label{svm} \bar\sigma_V(x) & =  \frac{1}{x+\bar m^2}\, \left[ 1 - {\cal F}(2
(x+\bar m^2))\right]\,,
\end{align}
\end{subequations}}
\hspace*{-0.5\parindent}with $x=p^2/\lambda^2$, $\bar m$ = $m/\lambda$,
\begin{equation}
\label{defcalF}
{\cal F}(x)= \frac{1-\mbox{\rm e}^{-x}}{x}  \,,
\end{equation}
$\bar\sigma_S(x) = \lambda\,\sigma_S(p^2)$ and $\bar\sigma_V(x) =
\lambda^2\,\sigma_V(p^2)$.
The mass-scale, $\lambda=0.566\,$GeV, and
parameter values
\begin{equation}
\label{tableA}
\begin{array}{ccccc}
   \bar m& b_0 & b_1 & b_2 & b_3 \\\hline
   0.00897 & 0.131 & 2.90 & 0.603 & 0.185
\end{array}\;,
\end{equation}
associated with Eqs.\,\eqref{EqSSSV} were fixed in analyses of light-meson observables \cite{Burden:1995ve, Hecht:2000xa}.\footnote{$\epsilon=10^{-4}$ in Eq.\ (\ref{ssm}) acts solely to decouple the large- and intermediate-$p^2$ domains.}
The expressions in Eq.\,\eqref{EqSSSV} guarantee dressed-quark confinement through the violation of reflection positivity; see, \emph{e.g}., Ref.\,\cite[Sect.\,5]{Ding:2022ows}.

The dimensionless $u=d$ current-mass, $\bar m$ in Eq.\,(\ref{tableA}), corresponds to
$m_q=5.08\,{\rm MeV}$ 
and the propagator yields the following Euclidean constituent-quark mass, defined by solving $p^2=M^2(p^2)$:
$M_q^E = 0.33\,{\rm GeV}$.
The ratio $M_q^E/m_q = 65$ is one expression of dynamical chiral symmetry breaking (DCSB), a corollary of EHM, in the parametrisation of $S(p)$.  It highlights the infrared enhancement of the dressed-quark mass function.

The dressed-quark mass function produced by \linebreak Eqs.\,\eqref{EqSSSV}\,--\,\eqref{tableA} is drawn in Ref.\,\cite[Fig.\,13]{Chen:2021guo}.  The image shows that although the algebraic representation is simple and was introduced long ago \cite{Burden:1995ve, Hecht:2000xa}, before practicable numerical solutions of QCD's gap equation were commonly obtained, the parametrisation is a sound representation of contemporary numerical results.

\subsection{Correlation amplitudes}
\label{subappendixdqbsa}
Following almost forty years of study, it has become clear that a realistic description of the nucleon can be achieved if, and only if, one includes the two positive-parity diquark correlations, \emph{viz}.\ isoscalar-scalar ($0^+$) and isovector-axial\-vector ($1^+$) diquarks.  Negative parity diquarks play practically no role in the positive parity nucleon \cite{Barabanov:2020jvn}.  Baryon structure models that ignore axialvector diquarks are incompatible with existing data; see, \emph{e.g}., Refs.\,\cite{Cui:2021gzg, Chen:2021guo, Chen:2023zhh}.

Regarding $0^+$, $1^+$ diquarks, the dominant structures in their correlation amplitudes are:
{\allowdisplaybreaks
\begin{subequations}
\label{qqBSAs}
\begin{align}
\label{scBSAs}
\Gamma^{0}(k;K) & = g_{0^+} \, \gamma_5 C\, t^0_f \,\vec{H}_c \,{\mathpzc F}(k^2/\omega_{0^+}^2) \,, \\
\label{axBSAs}
{\Gamma}_\mu^{1}(k;K)
    & = i g_{1^+} \, \gamma_\mu C \, \vec{t}_f\, \vec{H}_c \,{\mathpzc F}(k^2/\omega_{1^+}^2)\,.
\end{align}
\end{subequations}
Here,
$K$ is the correlation total momentum,
$k$ is a two-body relative momentum, ${\mathpzc F}$ is the function in Eq.\,\eqref{defcalF}, $\omega_{J^P}$ is a size parameter,
and $g_{J^P}$ is a coupling into the channel, which is fixed by normalisation -- see Eq.\,\eqref{CanNorm} below;
$C=\gamma_2\gamma_4$ is the charge-conjugation matrix;
$t^0$ and $\vec{t}=(t^1,t^2,t^3)$ introduce the flavour structure:
\begin{subequations}
\label{dqf}
\begin{align}
t^0_f &= \tfrac{i}{\surd 2}\tau^2\,,	\\
t^1_f &= \tfrac{1}{2}(\tau^0+\tau^3)\,,\\
t^2_f &= \tfrac{1}{\surd 2}\tau^1\,,\\
t^3_f &= \tfrac{1}{2}(\tau^0-\tau^3)\,,
\end{align}
\end{subequations}
$\tau^0=\,$diag$[1,1]$, $\{\tau^j,j=1,2,3\}$ are the Pauli matrices;
and $\vec{H}_c = \{i\lambda_c^7, -i\lambda_c^5,i\lambda_c^2\}$ express the diquarks' colour antitriplet character, where $\{\lambda_c^k,k=1,\ldots,8\}$ are the Gell-Mann matrices.

The amplitudes in Eqs.\,\eqref{qqBSAs} are nor\-malised:
\begin{subequations}
\label{CanNorm}
\begin{align}
2 K_\mu & = \left. \frac{\partial }{\partial Q_\mu} \,  \Pi(K;Q)\right|_{Q=K}^{K^2 = - m_{J^P}^2},\\
\nonumber
\Pi(K;Q) & = {\rm tr_{CDF}} \int \frac{d^4 k}{(2\pi)^4} \bar \Gamma(k;-K) S(k+Q/2) \\
& \quad \times \Gamma(k;K) S^{\rm T}(-k+Q/2)\,, \label{eqPi}
\end{align}
\end{subequations}
where $\bar\Gamma(k;K) = C^\dagger \Gamma(-k;K)^{\rm T} C$, with $(\cdot)^{\rm T}$ denoting matrix transpose.  When the amplitudes involved carry Lorentz indices, $\mu$, $\nu$, the left-hand-side of Eq.\,\eqref{eqPi} also includes a factor $\delta_{\mu\nu}$.
The coupling strength in each channel, $g_{J^P}$ in Eq.\,\eqref{qqBSAs}, is fixed by the associated value of $\omega_{J^P}$ via Eq.\,\eqref{CanNorm}.
}

\subsection{Diquark propagators and couplings}
\label{subappendixdqprop}
A propagator is connected with each quark + quark correlation in Fig.\,\ref{FigFaddeev}:
{\allowdisplaybreaks
\begin{subequations}
\label{Eqqqprop}
\begin{align}
\Delta^{0}(K^2) & = \frac{1}{m_{0^+}^2} \, {\mathpzc F}(K^2/\omega_{0^+}^2) \,,\\
\Delta^{1}_{\mu\nu}(K) & = \left[ \delta_{\mu\nu} + \frac{K_\mu K_\nu}{m_{1^+}^2} \right]
\Delta^{1}(K^2) \,,\\
\Delta^{1}(K^2) & =  \frac{1}{m_{1^+}^2} \, {\mathpzc F}(K^2/\omega_{1^+}^2)\,.
\end{align}
\end{subequations}
These representations ensure that each diquark is confined within the given baryon: whilst the propagators are free-particle-like at spacelike momenta, they are pole-free on the timelike axis.  This is enough to ensure violation of reflection positivity, hence, confinement.

The following values are commonly used for the masses in Eq.\,\eqref{Eqqqprop} \cite{Chen:2017pse, Cui:2020rmu, Chen:2021guo, Liu:2022ndb, Liu:2022nku}:
\begin{subequations}
\label{dqmasses}
\begin{align}
m_{[ud]_{0^+}} &= 0.79\,\mbox{GeV}\,,\\
m_{\{uu\}_{1^+}} & = m_{\{ud\}_{1^+}} = m_{\{dd\}_{1^+}} = 0.89\,\mbox{GeV}\,.
\end{align}
\end{subequations}
The widths in Eq.\,\eqref{Eqqqprop} are related to these masses via \cite{Segovia:2014aza}
\begin{equation}
\label{omegamass}
\omega_{J^P}^2 = \tfrac{1}{2} m_{J^P}^2 \,.
\end{equation}
This correspondence accentuates free-particle-like propagation characteristics for diquark correlations within the nucleon and also serves to reduce the number of parameters.

Working with Eqs.\,\eqref{qqBSAs}, \eqref{CanNorm}, \eqref{dqmasses}, \eqref{omegamass}, one finds
\begin{equation}
g_{0^+} = 14.94\,,\,\,\,\,\,
g_{1^+} = 12.98\,.
\end{equation}
Since it is the coupling-squared which appears in the Faddeev kernels, $0^+$ diquarks dominate the Faddeev amplitudes of $J=1/2^+$ baryons; but $1^+$ diquarks also play a measurable role because $g^2_{1^+}/g^2_{0^+} \simeq 0.75$.  These outcomes are visible, \emph{e.g}., in Ref.\,\cite[Fig.\,2]{Liu:2022ndb}.

\subsection{Faddeev amplitudes}
Having defined all elements in the kernel, it is possible to solve the Faddeev equation, Fig.\,\ref{FigFaddeev}, and obtain both the nucleon mass-squared, $m_N^2$, and bound-state amplitude.
As for all baryons, the nucleon is described by the sum of three amplitudes
\begin{align}
\Psi^N & = \psi^N_1 + \psi^N_2 + \psi^N_3\,,
\end{align}
where the subscript identifies the bystander quark, \emph{i.e}. the quark that is not participating in a diquark correlation.  $\psi^N_{1,2}$ are obtained from $\psi^N_3=:\psi^N$ by cyclic permutations of all quark labels.

Concerning the nucleon,
{\allowdisplaybreaks
\begin{align}
\nonumber
& \psi^N(p_i,\alpha_i,\sigma_i) \\
\nonumber
 = & [\Gamma^{0^+}(k;K)]^{\alpha_1 \alpha_2}_{\sigma_1 \sigma_2} \, \Delta^{0^+}(K) \,[\Psi^{0^+}(\ell;P) u(P)]^{\alpha_3}_{\sigma_3} \\
 & + [{\Gamma}^{j1^+}_\mu(k;K)]^{\alpha_1 \alpha_2}_{\sigma_1 \sigma_2} \, \Delta^{1^+ }_{\mu\nu}(K) \, [{\Psi}_{\nu }^{j1^+}(\ell;P) u(P)]^{\alpha_3}_{\sigma_3} \,,
\label{FaddeevAmp}
\end{align}
where
$(p_i,\alpha_i,\sigma_i)$ are the momentum, spin and isospin labels of the quarks comprising the bound state;
$P=p_1 + p_2 + p_3=p_d+p_q$ is the baryon's total momentum;
$k=(p_1-p_2)/2$, $K=p_1+p_2=p_d$, $\ell = (-K + 2 p_3)/3$;
the $j$ sum runs over the axialvector diquarks in the given bound state (Eq.\,\eqref{dqf} -- $1,2$ for proton and $2,3$ for neutron);
and $u(P)$ is a Euclidean spinor (see Ref.\,\cite{Segovia:2014aza}, Appendix\,B for details).

The hitherto undefined terms in Eq.\,\eqref{FaddeevAmp} are the following Dirac-matrix-valued functions:
\begin{subequations}
\label{8sa}
\begin{align}
\Psi^{0^+}(\ell;P) & = \sum_{k=1}^2 {\mathpzc s}_{k}(\ell^2,\ell\cdot P)\,  {\mathpzc S}^k(\ell;P) \,,
\\
{\Psi}_{\nu}^{j1^+}(\ell;P)  & = \sum_{k=1}^6 {\mathpzc a}_{k}^j(\ell^2,\ell\cdot P)\, \gamma_5 {\mathpzc A}^k_\nu(\ell;P) \,,
\end{align}
\end{subequations}
where
\begin{align}
\label{diracbasis}
\nonumber
{\mathpzc S}^1 & = {\mathbf I}_{\rm D} \,,\;
{\mathpzc S}^2  = i \gamma\cdot\hat\ell - \hat\ell \cdot\hat P {\mathbf I}_{\rm D}\,, \\
{\mathpzc A}^1_\nu & =  \gamma\cdot\ell^\perp \hat P_\nu\,,\;
{\mathpzc A}^2_\nu  = - i \hat P_\nu {\mathbf I}_{\rm D}\,,\;
{\mathpzc A}^3_\nu  = \gamma\cdot\hat\ell^\perp \hat\ell^\perp_\nu\,, \\
\nonumber
{\mathpzc A}^4_\nu & = i\hat \ell_\nu^\perp {\mathbf I}_{\rm D}\,,\;
{\mathpzc A}^5_\nu = \gamma_\nu^\perp - {\mathpzc A}^3_\nu\,,\;
{\mathpzc A}^6_\nu = i \gamma_\nu^\perp \gamma\cdot\hat\ell^\perp - {\mathpzc A}^4_\nu\,,
\end{align}
are the Dirac basis matrices, with $\hat\ell^2=1$, $\hat P^2 = -1$,
$\ell_\nu^\perp = \hat\ell_\nu +\hat\ell\cdot\hat P \hat P_\nu$,
$\gamma_\nu^\perp = \gamma_\nu +\gamma\cdot\hat P \hat P_\nu$;
}

\subsection{Nucleon Mass and Faddeev Amplitude}
In solving the Faddeev equation,
one projects the equation onto isospin $+1/2$ or $-1/2$ for proton or neutron,
then exploits a positive energy projection operator,
\begin{equation}
\label{Lplus} \Lambda_+(P):= \tfrac{1}{2 m_N}\,\sum_{s=\pm} \, u(P,s) \, \bar
u(P,s) = \tfrac{ -i \gamma\cdot P + m_N}{2m_N},
\end{equation}
in order to arrive at an integral equation for the eight scalar functions $\{{\mathpzc s}_{1,2}(\ell^2,\ell\cdot P)\}$, $\{{\mathpzc a}_{1,\ldots, 6}(\ell^2,\ell\cdot P)\}$ that describe the relative quark + diquark momentum correlation inside the nucleon.  (Isospin symmetry entails, \emph{e.g}., $\surd 2 a_k^{2} = - a_k^{1}$.  Details are provided elsewhere \cite[Sect.\,2]{Alkofer:2004yf}.)
It is worth noting that the direct three-body approach involves $128$ functions of this type; so, the quark + diquark approximation delivers a considerable simplification.

Using the inputs described above, the Faddeev equation yields
\begin{equation}
m_N = 1.18\,{\rm GeV}.
\end{equation}
This value is deliberately large because Fig.\,\ref{FigFaddeev} generates what is typically described as the nucleon's dressed-quark core.
The complete (physical) nucleon is obtained by adding resonant (meson-cloud) contributions to the Faddeev kernel.
Such effects generate a reduction of $\approx 0.2\,\text{GeV}$ in the physical nucleon mass \cite{Hecht:2002ej, Sanchis-Alepuz:2014wea}.
Whilst their impact on nucleon structure can be estimated using dynamical coupled-channels models \cite{Aznauryan:2012ba, Burkert:2017djo}, that is beyond the scope of contemporary Faddeev equation analyses.
For this reason, we follow the common practice of expressing form factors in terms of $x=Q^2/m_N^2$.

\section{Nucleon Electromagnetic Current}
\label{SecCurrent}
The nucleon electromagnetic interaction current can be written as follows:
\begin{subequations}
\label{current}
\begin{align}
\label{Jnucleon}
J_\mu & (P_f,P_i) = ie\,\Lambda_+(P_f)\, \Lambda_\mu(Q,P) \,\Lambda_+(P_i)\,, \\
& = i e \,\Lambda_+(P_f)\,\left[ \gamma_\mu F_1(Q^2) \right. \nonumber \\
& \qquad + \left.
\tfrac{1}{2m_N}\, \sigma_{\mu\nu}\,Q_\nu\,F_2(Q^2)\right] \,\Lambda_+(P_i)\,,
\label{JnucleonB}
\end{align}
\end{subequations}
where $P_{i,f}$ are the momenta of the incoming, outgoing nucleon, $Q= P_f - P_i$, and $F_1$ and $F_2$ are, respectively, the Dirac and Pauli form factors.
One may equally express the current in terms of the proton charge and magnetisation distributions ($\tau = Q^2/[4 m_N^2]=x/4$) \cite{Sachs:1962zzc}:
\begin{subequations}
\label{EqSachs}
\begin{align}
\label{GEpeq}
G_E(Q^2)  & =  F_1(Q^2) - \tau F_2(Q^2)\,,\; \\
G_M(Q^2)  & =  F_1(Q^2) + F_2(Q^2)\,.
\end{align}
\end{subequations}

As usual, we construct the photon-nucleon vertex, $\Lambda_\mu$ in Eq.\,\eqref{Jnucleon},  following the systematic procedure of Ref.\,\cite{Oettel:1999gc}.  That scheme has the merit of automatically providing a conserved current for on-shell nucleons described by Faddeev amplitudes of the type we have calculated.
Canonical normalisation of the nucleon Faddeev amplitude is implemented by fixing $F_1(Q^2=0)=1$ for the proton; then current construction ensures \linebreak $F_1^{\rm neutron}(Q^2=0)=0$.

\begin{figure}[t]
\centerline{\includegraphics[clip, width=0.49\textwidth]{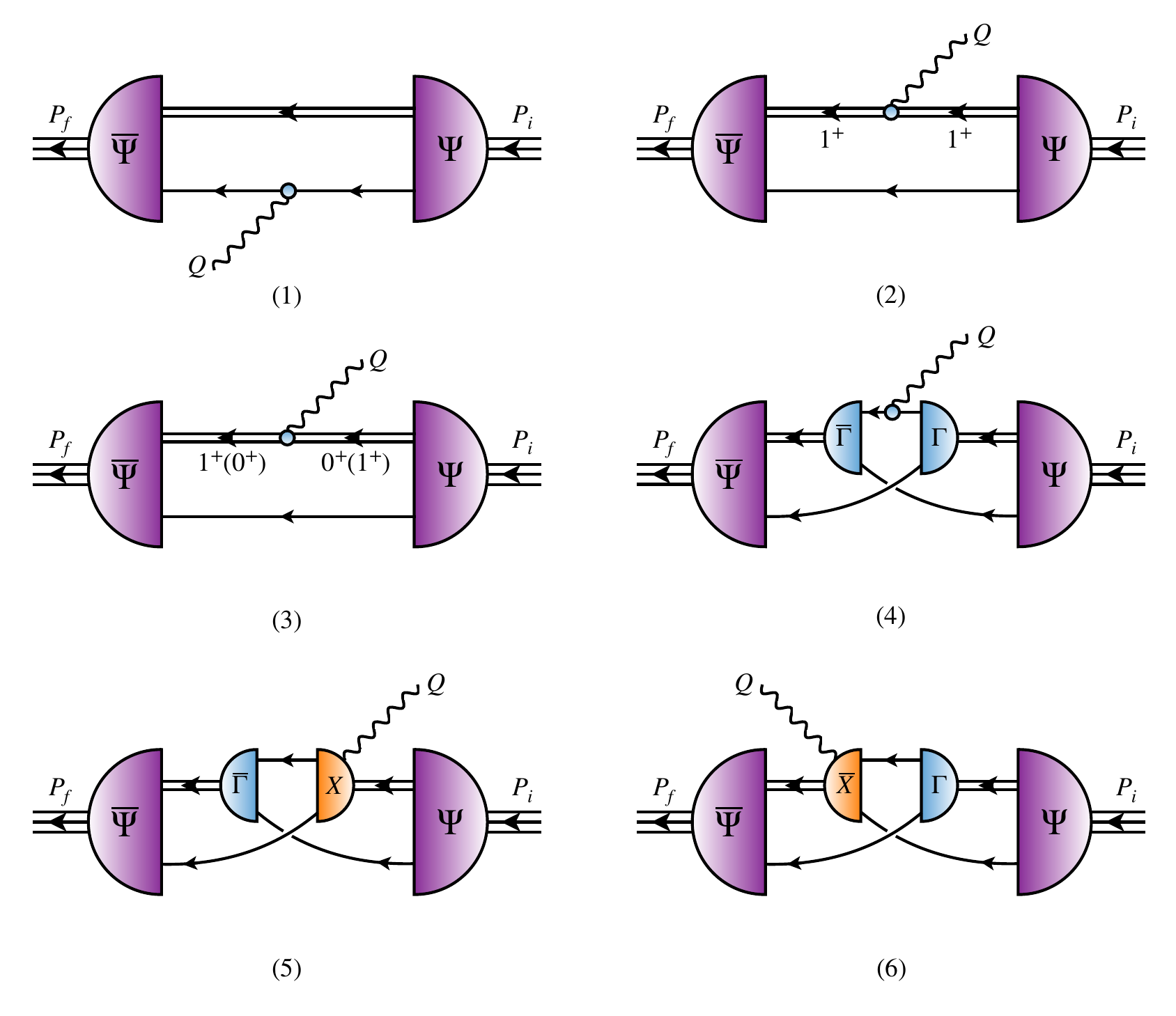}}
\caption{\label{figcurrent} The collection of electromagnetic interaction diagrams that ensures a conserved current for on-shell nucleons described by the Faddeev amplitude, $\Psi$, calculated as described in Sect.\,\ref{SecFaddeev}.
Legend.
\emph{single line}, dressed-quark propagator; \emph{undulating line}, electromagnetic probe; $\Gamma$,  diquark correlation amplitude; \emph{double line}, diquark propagator; and $X$, seagull terms.
}
\end{figure}

The interaction current has six terms, which are depicted in Fig.\,\ref{figcurrent} and represent the following expression:
\begin{align}
J_\mu & (P_f,P_i) =
\int \frac{d^4 p}{(2\pi)^4} \frac{d^4k}{(2\pi)^4} \nonumber\\
& \qquad
\sum_{i=1}^6 \, \bar\Psi(p,-P_f)  \Lambda_\mu^i(p,P_f;k,P_i) \Psi(k;P_i)\,,
\label{NCurrent}
\end{align}
where $\bar \Psi(p,-P) = C^\dagger \Psi(-p,-P)^{\rm T} C $.
In practical calculations, we work in the Breit frame: $P_{f/i} = P \pm Q/2$, $P=(0,0,0,i E)$, $E^2= M^2 + Q^2/4$.

It is worth noting that our approach is formulated at the hadron scale, $\zeta_{\cal H}$, whereat all properties of a given hadron are carried by its valence degrees of freedom.
In the case under consideration, this means that partonic glue and sea-quark contributions to measurable form factors are sublimated into the dressed valence quarks at $\zeta_{\cal H}$.
As merely a definition of dressing, such is the content of any complete QCD Schwinger function; consider, \textit{e.g}., a diagrammatic expansion of the quark gap equation.

The existence and character of $\zeta_{\cal H}$ are discussed elsewhere \cite{Yin:2023dbw}; and an appreciation of its role has proved useful in developing predictions for a diverse array of parton distribution and fragmentation functions \cite{Raya:2021zrz, dePaula:2022pcb, Ding:2022ows, Raya:2024ejx, Yao:2024ixu, Xu:2024nzp, Xing:2025eip}.
Since the measurable nucleon elastic form factors considered herein are scale invariant, formulating the calculations at $\zeta_{\cal H}$ has no material impact on them, but it does have implications for any separations/decompositions of these form factors into contributions from various parton species.
Examples are provided in Refs.\,\cite{Raya:2021zrz, Yao:2024ixu}.

Our construction of the individual contributions in Eq.\,\eqref{NCurrent} differs from that in Refs.\,\cite{Cui:2020rmu, Segovia:2014aza}, so we now detail each in turn.

\subsection{Diagram 1}
\label{Diag1}
This expresses the photon coupling directly to the bystander quark:
\begin{align}
 \Lambda_\mu^1& (p,P_f;k,P_i)  =
 i {\mathpzc Q} \chi_\mu(p_q,k_q) \nonumber \\
 & \quad \times \left[ \Delta^{0}(k_d) + \Delta^1(k_d)\right] (2\pi)^4 \delta^4(p-k-\hat\eta Q)\,,
 \label{CD1}
\end{align}
where, for notational simplicity, we have suppressed the subscripts on $\Delta^1$, Eq.\,\eqref{Eqqqprop}, which connect to partners in the Faddeev amplitude;
${\mathpzc Q} = {\rm diag}[e_u,e_d]$ is the quark electric charge matrix; $\chi_\mu(p_q,k_q)$ is the unamputated dressed-photon-quark vertex; and the momenta are ($\eta+\hat \eta =1$):
\begin{equation}
\begin{array}{ll}
k_q = \eta P_i + k \,, & p_q = \eta P_f + p \,, \\
k_d = \hat\eta P_i - k \,, & p_d = \hat\eta P_f - p \,.
\end{array}
\end{equation}

We use $\eta = 1/3$ throughout, so that the incoming photon momentum is shared equally between the dressed valence quarks in the nucleon.  This simplifies numerical analyses and accelerates convergence of calculations.  Naturally, since our approach is Poincar\'e covariant, the results are independent of $\eta$ so long as the numerical methods preserve that covariance.

Current conservation requires that the photon-quark vertex satisfy a Ward-Green-Takahashi identity:
\begin{align}
\label{eqWGTI}
i (\ell_1 - \ell_2)_\mu \chi_\mu(\ell_1,\ell_2) & = S(\ell_2) - S(\ell_1)\,.
\end{align}
Since the dressed quark propagator has been paramet\-rised, we generalise Ref.\,\cite{Ding:2018xwy} and use the following \linebreak \emph{Ansatz} for the vertex, expressed in terms of quantities already specified:
\begin{align}
\chi_\mu(\ell_1,\ell_2) & = \gamma_\mu \Sigma_{\sigma_V}
+ 2 \ell_\mu [ \gamma\cdot \ell \Delta_{\sigma_V}+ i \Delta_{\sigma_S}]  \nonumber \\
& \quad +  \tfrac{1}{2} [s_1-\bar s_1][\gamma\cdot \check{\ell} \gamma_\mu \gamma\cdot \ell  - \gamma\cdot\ell \gamma_\mu \gamma\cdot \check{\ell}] \Delta_{\sigma_V} \nonumber \\
& \quad - [s_2-\bar s_2]\sigma_{\mu\nu} \check\ell_\nu \Delta_{\sigma_S} \nonumber \\
& \quad + \tfrac{i}{2} [1+s_3] \gamma\cdot\check{\ell}\sigma_{\mu\nu}\check{\ell}_\nu \Delta_{\sigma_V}\,,
\label{chiquark}
\end{align}
where
\begin{subequations}
\begin{align}
\Sigma_F & = \tfrac{1}{2}[ F(\ell_1^2) + F(\ell_2^2)]\,, \\
\Delta_F & = [ F(\ell_1^2) - F(\ell_2^2)]/[\ell_1^2-\ell_2^2]\,,
\end{align}
\end{subequations}
$\ell = [\ell_1+\ell_2]/2$, $\check\ell = \ell_1 - \ell_2$, and, $\bar s_{1,2} = 1-s_{1,2}$
\begin{subequations}
\begin{align}
s_i & = a_i + b_i \exp[-\tfrac{1}{4}{\mathpzc E}(\check\ell)/M_q^E]\,, \\
{\mathpzc E}(\check\ell)/M_q^E & = (1+\check\ell^2/[2 M_q^E]^2)^{1/2} - 1\,.
\end{align}
\end{subequations}
Here, the parameters $a_{1,2,3}$, $b_{1,2,3}$ characterise the transverse part of the vertex \emph{Ansatz}, which is not constrained by Eq.\,\eqref{eqWGTI}.
They are used to obtain agreement with the three-body nucleon form factor results in Ref.\,\cite{Yao:2024uej}, which employed a photon-quark vertex whose eleven distinct characterising functions were obtained by solving the relevant inhomogeneous vertex equation.

In Refs.\,\cite{Cui:2020rmu, Segovia:2014aza}, it was the amputated vertex for which an \emph{Ansatz} was introduced. That approach has the potential for introducing singularities into \linebreak inte\-grands which appear in form factor calculations.  To avoid that, we have adopted the Ref.\,\cite{Ding:2018xwy} procedure herein, \emph{i.e}., parametrised the unamputated vertex.

\subsection{Diagram 2}
\label{Diag2}
In this case, the photon probes the scalar or axialvector diquark:
\begin{align}
\Lambda_\mu^2 (p,P_f; & k,P_i)  = [\chi_\mu^{qq}(p_d,k_d)] \nonumber \\
& \quad \times S(k_q) (2\pi)^4 \delta^4(p-k-\eta Q)\,,
\end{align}
where
$[\chi_\mu^{qq}(p_d,k_d)] = {\rm diag}[e_{0} \chi_\mu^{0},e_{1}^i \chi_\mu^1]$, $e_0 = 1/3$, $\{e_1^i\} = \{4/3, 1/3, -2/3\}$ and the $\chi$ functions that appear express the photon-diquark form factors.

Regarding the scalar diquark, we employ the following Ward-Green-Takahashi identity preserving \emph{Ansatz}:
\begin{align}
\label{Gamma0plus}
\chi_\mu^0(\ell_1,\ell_2) & = -2 \ell_\mu \Delta_{\Delta^0}(\ell_1^2,\ell_2^2) F_{\rm J=sc}(\check\ell^2)\,,
\end{align}
where
\begin{equation}
F_{\rm J}(\check\ell^2) = 1/[1+r_{\rm J}^2 \check\ell^2/6]\,.
\label{qqradii}
\end{equation}
This is practically equivalent to the form used in \linebreak Refs.\,\cite{Cui:2020rmu, Segovia:2014aza}.  Therein, however, repeating an error in Ref.\,\cite{Cloet:2008re}, the diquark radius was set to zero: $r_{\rm sc}=0$.  We use a nonzero value herein, chosen so as to recover selected three-body form factor results  in Ref.\,\cite{Yao:2024uej}.

We introduce the following \emph{Ansatz} for the unamputated photon--axialvector-diquark vertex:
\begin{align}
\chi_{\mu\alpha\beta}^1 & (\ell_1,\ell_2) = - F_{\rm av}(\check\ell^2) \big\{ 2 \delta_{\alpha\beta} \ell_\mu \Delta_{\Delta^1}(\ell_1^2,\ell_2^2) \nonumber \\
& \quad + \tfrac{1}{m_{1^+}^2} \big[\delta_{\mu\alpha} \ell_{1\beta} \Delta^1(\ell_1^2)
+ \delta_{\mu\beta}\ell_{2\alpha} \Delta^1(\ell_2^2) \big]  \nonumber \\
& \quad + \tfrac{2}{m_{1^+}^2} \ell_\mu \ell_{2\alpha}\ell_{1\beta} \Delta_{\Delta^1}(\ell_1^2,\ell_2^2) \}
\nonumber \\
& \quad - \kappa [ \check\ell_\beta \delta_{\mu\alpha} - \check\ell_\alpha \delta_{\mu\beta}]
\Delta_{\Delta^1}(\ell_1^2,\ell_2^2)\,, \label{chiav}
\end{align}
where $\kappa$ is a parameter that serves to characterise the strength of a possible axialvector diquark anomalous magnetic interaction \cite{Oettel:2000jj}.
In our calculations, we find that it has little impact; hence, hereafter, we set $\kappa \equiv 0$.

The vertex in Eq.\,\eqref{chiav} satisfies a Ward-Green-Taka\-hashi identity:
\begin{equation}
\label{avwgti}
 \check\ell_\mu \chi_{\mu\alpha\beta}^1(\ell_1,\ell_2)  = F_{\rm av}(\check\ell^2)
  \big[\Delta_{\alpha\beta}^1(\ell_2) - \Delta_{\alpha\beta}^1(\ell_1)\big]\,.
\end{equation}
It improves upon the \emph{Ansatz} used in Refs.\,\cite{Cui:2020rmu, Segovia:2014aza, Cloet:2008re}, which failed to account for the fact that the axialvector diquark correlation amplitude in Eq.\,\eqref{axBSAs} is not itself transverse.

For an on-shell (transverse) axialvector diquark, the amputated elastic photon--axialvector-diquark vertex \linebreak can be written $(k^2 T_{\mu\nu}(k) = k^2\delta_{\mu\nu} - k_\mu k_\nu)$ \cite{Cloet:2008re}:
\begin{subequations}
\label{ampavvtx}
\begin{align}
\Gamma_{\mu\alpha\beta}^1(\ell_1,\ell_2) & =
-\sum_{i=1,2,3} \Gamma_{\mu\alpha\beta}^{[i]}(\ell_1,\ell_2)\,, \\
\Gamma_{\mu\alpha\beta}^{[1]}(\ell_1,\ell_2) & =
2 \ell_\mu T_{\alpha\rho}(\ell_1) T_{\rho\beta}(\ell_2) F_1(\ell_1^2,\ell_2^2) \,, \\
\Gamma_{\mu\alpha\beta}^{[2]}(\ell_1,\ell_2) & =
\big[ T_{\mu\alpha}(\ell_1)\, T_{\beta\rho}(\ell_2) \, \ell_{1 \rho} \nonumber \\
& \quad + T_{\mu\beta}(\ell_2) \, T_{\alpha\rho}(\ell_1) \, \ell_{2\rho} \big] F_{2}(\ell_1^2,\ell_2^2) \,, \\
\Gamma^{\rm [3]}_{\mu\alpha\beta}(\ell_1,\ell_2)
&= -\frac{1}{ m_{1^+}^2}\, \ell_\mu \, T_{\alpha\rho}(\ell_1)\, \ell_{2 \rho} \nonumber \\
& \quad \times T_{\beta\lambda}(\ell_2)\, \ell_{1 \lambda}\; F_{3}(\ell_1^2,\ell_2^2) \,.
\end{align}
\end{subequations}
Axialvector diquark electric, magnetic, and quadrupole form factors may now be expressed as follows:
\begin{subequations}
\label{avqqEMQ}
\begin{align}
G_{\cal E}^{1} & = F_1 + \tfrac{2}{3}\, \tau_{1^+}\, G_{\cal Q}^{1} \,, \; \tau_{1} = \frac{\check\ell^2}{ 4 \,m_{1^{+}}^{2}}\,, \\
G_{\cal M}^{1} & = - F_2 \, \\
G_{\cal Q}^{1} & = F_1 + F_2 + (1 + \tau_{1^+}) F_3\,.
\end{align}
\end{subequations}

In the context of our analysis, one can infer these axialvector diquark form factors from Eq.\,\eqref{chiav} by forming the contraction
\begin{equation}
  \left.
  \Pi^1(\ell_1^2) \, T_{\alpha\rho}(\ell_1) \, \chi_{\mu\rho\sigma}^1(\ell_1,\ell_2) \, T_{\sigma\beta}(\ell_2) \Pi^1(\ell_2^2) \, \right|_{\ell_1^2=\ell_2^2}\,,
\end{equation}
where $\Pi^1 = 1/\Delta^1$, and comparing the result with \linebreak Eq.\,\eqref{ampavvtx}:
\begin{subequations}
\label{avF123}
\begin{align}
F_1 & = F_{\rm av}(\check\ell^2) \frac{d}{d\ell^2} \Pi^1(\ell^2)  \,, \\
%
F_2 & = - F_{\rm av}(\check\ell^2) \tfrac{1}{m_1^2}\Pi^1(\ell^2) - \kappa \frac{d}{d\ell^2} \Pi^1(\ell^2)\,, \\
%
F_3 & = -2 F_{\rm av}(\check\ell^2) \frac{d}{d\ell^2} \Pi^1(\ell^2)  \,.
\end{align}
\end{subequations}
Using Eqs.\,\eqref{avqqEMQ}, \eqref{avF123}, one arrives at results for axialvector diquark static electromagnetic properties, \emph{viz}.\ charge, magnetic moment, and quadrupole moment:
\begin{subequations}
\begin{align}
\left. G_{\cal E}^{1}(\ell^2,\ell^2)\right|_{\ell^2=0} & = \left. \frac{d}{d\ell^2} \Pi^1(\ell^2) \right|_{\ell^2=0} = 1\,, \\
\left. G_{\cal M}^{1}(\ell^2,\ell^2)\right|_{\ell^2=0}& = 1 + \kappa =: \mu^1\,, \\
\left. G_{\cal Q}^{1}(\ell^2,\ell^2)\right|_{\ell^2=0}& = -(2+\kappa) =: -{\cal Q}^1\,.
\end{align}
\end{subequations}
Recall that we set $\kappa\equiv 0$ in our calculations.  It is included here simply for illustrative purposes.

\subsection{Diagram 3}
\label{Diag3}
This image depicts the form factor contribution from electromagnetically induced $0^+ \leftrightarrow 1^+$ diquark transitions:
\begin{align}
\Lambda_\mu^{3\{i\neq j\}}(p,P_f; & k,P_i) =
\Delta^i(p_d) \big[ \chi_\mu^{qq}(p_d,k_d)\big]^{\{ij\}} \Delta^j(p_d) \nonumber \\
& \quad \times  S(k_q) (2\pi)^4 \delta^4(p-k+\eta Q)\,,
\end{align}
with
$\big[ \chi_\mu^{qq}(p_d,k_d)\big]^{\{01\}} = \chi^{01}_{\mu\alpha}(p_d,k_d)$,
$\big[ \chi_\mu^{qq}(p_d,k_d)\big]^{\{10\}} = \chi^{10}_{\mu\alpha}(p_d,k_d)$,
\begin{align}
\chi^{01}_{\mu\alpha}(\ell_1,\ell_2) & = - \chi^{10}_{\mu\alpha}(\ell_1,\ell_2) \nonumber \\
& = \tfrac{i}{M} \kappa_{01}\varepsilon_{\mu\alpha\rho\lambda} \ell_{1\rho}\ell_{2\lambda} F_{\rm av}(\hat\ell^2)\,.
\end{align}
The transition strength is not varied herein.  Instead, working from Ref.\,\cite{Oettel:2000jj}, we set $\kappa_{01} = 2.64$, which is roughly twice the size of that used in Refs.\,\cite{Cui:2020rmu, Segovia:2014aza}.
%
Otherwise, this current contribution is the same as therein.

\subsection{Diagram 4}
\label{Diag4}
This two-loop diagram is the form factor contribution produced by the photon as it couples to the quark that is exchanged when one diquark breaks up and another is formed.  It is expressed as follows:
\begin{align}
\label{B3}
\Lambda_{\mu}^{4\{ij\}}& (p,P_f; k,P_i) = -\frac{1}{2} S(p_{q}) \Delta^{i}(p_{d})
\Gamma^{j}(p_1 ; k_{d}) \nonumber \\
& \times i{\mathpzc Q} \chi_\mu(q^\prime,q)^{\rm T}\bar{\Gamma}^{i{\rm T}}(p'_2 ; -p_{d})
\Delta^j(k_{d}) S(k_{q})\,,
\end{align}
where the photon-quark vertex, $\chi_\mu$, appeared in Eq.\,\eqref{CD1} and the charge conjugated correlation amplitude is defined after Eq.\,\eqref{CanNorm}.  The momenta are:
\begin{eqnarray}
\begin{array}{lc@{\quad}l}
q = \hat{\eta}P_i-\eta P_f-p-k\,, & & q' = \hat{\eta}P_f-\eta P_i-p-k \,,\\
p_1 = (p_q-q)/2\,,& & p'_2 = (-k_q+q')/2 \,,\\
p'_1 = (p_q-q')/2\,, & & p_2 = (-k_q+q)/2 \,.
\end{array}
\end{eqnarray}
The full contribution ($4$ terms) is obtained by summing over the superscripts $\{i=0,1; j=0,1\}$.  When required, the axialvector diquark subscripts can readily be made explicit.

It is worth highlighting that quark exchange provides the attraction necessary in the Faddeev equation to bind the nucleon.  It also ensures that the Faddeev amplitude has the correct antisymmetry under the exchange of any two dressed-quarks.  This crucial feature is lacking in models with elementary (noncomposite) diquarks.

Our construction of Diagram~4 is precisely the same as that in Refs.\,\cite{Cui:2020rmu, Segovia:2014aza}.  Therein, however, following Ref.\,\cite{Cloet:2008re}, a multiplicative factor of $0.406973$ was included to damp the contribution.  (The same factor was used for Diagrams~5 and 6.)  We cannot find a justification for retaining that factor, so we have replaced it by unity.

\subsection{Diagrams~5 \& 6}
\label{X56}
These two-loop diagrams are typically called ``seagull'' terms.  They appear as partners of Diagram~4 and arise because binding in the nucleon Faddeev equation is effected by the exchange of an electrically charged dressed-quark between nonpointlike diquark correlations \cite{Oettel:1999gc}.
The explicit expression for their contribution to the nucleon form factors is
\begin{align}
\label{B5}
\Lambda_{\mu}^{56}& (p,P_f; k,P_i) =
\frac{1}{2} S(p_{q}) \Delta^{i}(p_{d}) \nonumber \\
& \quad \times \left[ X_{\mu}^{j}(p_1^\prime ; k_d) S^{\rm T}(q')
\bar{\Gamma}^{i{\rm T}}(p^\prime_2 ; - p_{d})
\right.
\nonumber\\
&
\left. \qquad -
\Gamma^{j}(p_1,k_{d}) S^{\rm T}(q)
\bar{X}_{\mu}^{i}(-p_2; - p_d)
\right] \nonumber \\
& \quad \times \Delta^{j}(k_{d}) S(k_{q})\,,
\end{align}
where the superscripts are summed and, when necessary, the axialvector diquark subscripts can readily be made explicit.

The novel elements here are the couplings of a photon to two dressed-quarks as they either separate from (Diagram~5) or combine to form (Diagram~6) a diquark correlation.
Consequently, they are components of the five point Schwinger function that describes the coupling of a photon to the quark + quark scattering kernel.
This Schwinger function could be calculated.
In fact, computations of analogous Schwinger functions relevant to meson observables exist; see, \emph{e.g}., Refs.\,\cite{Bicudo:2001jq, Cotanch:2002vj, Ding:2019lwe}.

However, such a calculation would provide relevant input only when a uniformly composed CSM truncation is employed to calculate each of the elements described hitherto.
Absent that, we instead employ an algebraic parametrisation \cite{Oettel:1999gc}, which for Diagram~5, \emph{viz}.\ the first term in the difference expressed by Eq.\,\eqref{B5}, reads
\begin{align}
\nonumber
& X^{J^P}_\mu (l;K)  = e_{\rm by}\,\frac{4 l_\mu- K_\mu}{4 l\cdot K - K^2}\,\left[\Gamma^{J^P}\!(l-K/2)-\Gamma^{J^P}\!(l)\right]\\
& \quad + e_{\rm ex}\,\frac{4 l_\mu+ K_\mu}{4 l\cdot K + K^2}\,\left[\Gamma^{J^P}\!(l+K/2)-\Gamma^{J^P}\!(l)\right], \label{X5}
\end{align}
with $l$ the relative momentum between the quarks in the initial diquark, $e_{\rm by}$ the electric charge of the quark which becomes the bystander, and $e_{\rm ex}$ the charge of the quark that is reabsorbed into the final diquark.
Diagram~6, expressed by the second term in Eq.\,\eqref{B5}, has
\begin{align}
\nonumber
& \bar X^{J^P}_\mu(l,K)  = e_{\rm by}\,\frac{4 l_\mu+ K_\mu}{4 l\cdot K + K^2}\,\left[\bar\Gamma^{J^P}\!(l+K/2)-\bar\Gamma^{J^P}\!(l)\right]\\
& \quad +e_{\rm ex}\,\frac{4 l_\mu-K_\mu}{4 l\cdot K - K^2}\,\left[\bar\Gamma^{J^P}\!(l-K/2)-\bar\Gamma^{J^P}\!(l)\right], \label{X6}
\end{align}
where $\bar\Gamma^{J^P}\!(\ell)$ is the charge-conjugated amplitude.

Plainly, the seagull terms are identically zero if the diquark correlation is represented by a momentum-inde\-pen\-dent Bethe-Salpeter-like amplitude.  This feature has been exploited in a variety of analyses whose aim was to expose the sensitivity of observables to the momentum dependence of QCD Schwinger functions; see, \emph{e.g}., Refs.\,\cite{Wilson:2011aa, Segovia:2013rca, Xu:2015kta, Raya:2021pyr}.

It is worth stressing that, like Eq.\,(\ref{Gamma0plus}), Eqs.\,(\ref{X5}) and (\ref{X6}) are simple forms, free of kinematic singularities and sufficient to ensure the nucleon-photon interaction vertex -- Fig.\,\ref{figcurrent} -- satisfies the relevant Ward-Green-Takahashi identity when the composite nucleon is obtained from the Faddeev equation sketched in Fig.\,\ref{FigFaddeev}.  Thus, they are sufficient to our needs, as they were in Refs.\,\cite{Cui:2020rmu, Segovia:2014aza}.   As noted above, herein, we replace by unity the multiplicative factor of $0.406973$ applied to these contributions in Refs.\,\cite{Cui:2020rmu, Segovia:2014aza}.

\section{Nucleon Electromagnetic Form Factors: Method of Calculation}
\label{Method}
We have introduced a refinement of the QCD-kindred model that has hitherto been used in numerous analyses of nucleon elastic and transition form factors; see, \emph{e.g}., Ref.\,\cite{Segovia:2014aza} and citations thereof.  The model has eight parameters:
six are used to deliver a flexible photon-quark vertex, Eq.\,\eqref{chiquark};
and there are scalar and axialvector diquark radii, Eq.\,\eqref{qqradii}.
We determine values for these parameters via a least-squares fit to ($x=Q^2/m_N^2$)
$G_E^p(x)$ on $x\in (0,4.5)$ and $\mu_p G_E^p(x)/G_M^p(x)$ on $x\in (2,4.5)$.
This procedure yields the values listed in Table~\ref{list9parameters}.

\begin{table}[t]
\caption{\label{list9parameters}
Parameters completing the definition of our photon-nucleon vertex: Eqs.\,\eqref{chiquark}, \eqref{qqradii}.}
\begin{center}
\begin{tabular*}
{\hsize}
{
c@{\extracolsep{0ptplus1fil}}
c@{\extracolsep{0ptplus1fil}}
c@{\extracolsep{0ptplus1fil}}
c@{\extracolsep{0ptplus1fil}}
c@{\extracolsep{0ptplus1fil}}
c@{\extracolsep{0ptplus1fil}}}\hline
$a_1\ $ & $a_2\ $ & $a_3\ $ & $b_1\ $ & $b_2\ $ & $b_3\ $ \\
$0.17\ $ & $1.47\ $ & $-0.93\ $ & $1.03\ $ & $-0.17\ $ & $1.61\ $
\\\hline
$r_{\rm sc}/{\rm fm}\ $ & $r_{\rm av}/{\rm fm}\ $ & &  &  & \\
$0.31\ $ & $ 1.05\ $ & & & & \\\hline
\end{tabular*}
\end{center}
\end{table}

In computing all form factors, we use the algorithm described in Ref.\,\cite[Appendix~D]{Segovia:2014aza}.
Regarding the Faddeev equation, ten Chebyshev polynomials are employed to express the $\ell\cdot P$ dependence of the Faddeev amplitudes, $\Psi^{0^+,1^+}(\ell;P)$ in Eq.\,\eqref{FaddeevAmp}.
Concerning the integrals depicted in Fig.\,\ref{figcurrent}: to compute Diagrams~1\,--\,3, 150 quadrature points are used for the momentum magnitude and 50 for both angles; and for Diagrams~4\,--\,6, $10^8$ Monte Carlo points.  These choices ensure stable results on $x\leq 6$.

On $x\gtrsim 6$, slowing convergence of the Chebyshev expansions of $\Psi^{0^+,1^+}(\ell;P)$, \emph{viz}.\ increasing noise and spurious oscillations, exacerbates inadequacies in the Monte Carlo approach to integrating functions with non-uni\-form sign.
Consequently, the direct (brute force) approach fails to deliver precise results for the form factors.
We overcome this difficulty by using the Schlessin\-ger point method (SPM) \cite{Schlessinger:1966zz, PhysRev.167.1411, Tripolt:2016cya}.

The SPM is grounded in analytic function theory.  It accurately reconstructs any function in the complex plane within a radius of convergence determined by that one of the function's branch points which lies closest to the real domain that provides the sample points.
Our implementation introduces a statistical element; so, the extrapolations come with an objective and reliable estimate of uncertainty.
Crucially, the SPM is free from practitioner-induced bias; thus, delivers objective analytic continuations and extrapolations.
In practice, the SPM has been blind-tested against numerous models and physically validated in applications that include
extraction of hadron and light nucleus radii from electron scattering \cite{Cui:2022fyr};
determination of resonance properties from scattering data \cite{Binosi:2022ydc};
searching for evidence of the odderon in high-energy elastic hadron + hadron scattering \cite{Cui:2022dcm};
and
calculation of meson and baryon electromagnetic and gravitational form factors \cite{Yao:2024drm, Yao:2024uej, Yao:2024ixu}.

\begin{table}[t]
\caption{\label{StaticResults}
Calculated results for $Q^2\simeq 0$ (static) properties of nucleon electromagnetic form factors.  Also listed, for comparison, results obtained using the \emph{ab initio} three-body approach \cite{Yao:2024uej} and experiment \cite[PDG]{ParticleDataGroup:2024cfk}.}
\begin{center}
\begin{tabular*}
{\hsize}
{
r@{\extracolsep{0ptplus1fil}}
|c@{\extracolsep{0ptplus1fil}}
c@{\extracolsep{0ptplus1fil}}
c@{\extracolsep{0ptplus1fil}}
c@{\extracolsep{0ptplus1fil}}
c@{\extracolsep{0ptplus1fil}}}\hline
 & $r_E^2/{\rm fm}^2\ $ & $r_E^2 M^2\ $ & $r_M^2/{\rm fm}^2\ $ & $r_M^2 M^2\ $ & $\mu\ $ \\ \hline
herein \; $p\ $ & $\phantom{-}0.67^2\ $ & $\phantom{-}4.02^2\ $& $0.67^2\ $ & $4.02^2\ $ & $\phantom{-}2.80\ $ \\
$n\ $ & $-0.29^2\ $ & $-1.73^2\ $& $0.72^2\ $ & $4.30^2\ $ & $-1.86\ $ \\
\hline
\cite{Yao:2024uej} \; $p\ $ & $\phantom{-}0.89^2\ $ & $\phantom{-}4.23^2\ $& $0.82^2\ $ & $3.91^2\ $ & $\phantom{-}2.23\ $ \\
$n\ $ & $-0.25^2\ $ & $-1.19^2\ $& $0.81^2\ $ & $3.87^2\ $ & $-1.33\ $ \\
\hline
\cite[PDG]{ParticleDataGroup:2024cfk} \; $p\ $ & $\phantom{-}0.84^2\ $ & $\phantom{-}4.00^2\ $& $0.85^2\ $ & $4.05^2\ $ & $\phantom{-}2.79\ $ \\
$n\ $ & $-0.34^2\ $ & $-1.62^2\ $& $0.86^2\ $ & $4.11^2\ $ & $-1.91\ $ \\
\hline
\end{tabular*}
\end{center}
\end{table}

\begin{figure*}[t]
\begin{minipage}[t]{\textwidth}
\begin{minipage}[t]{0.48\textwidth}
\includegraphics[width=0.95\textwidth]{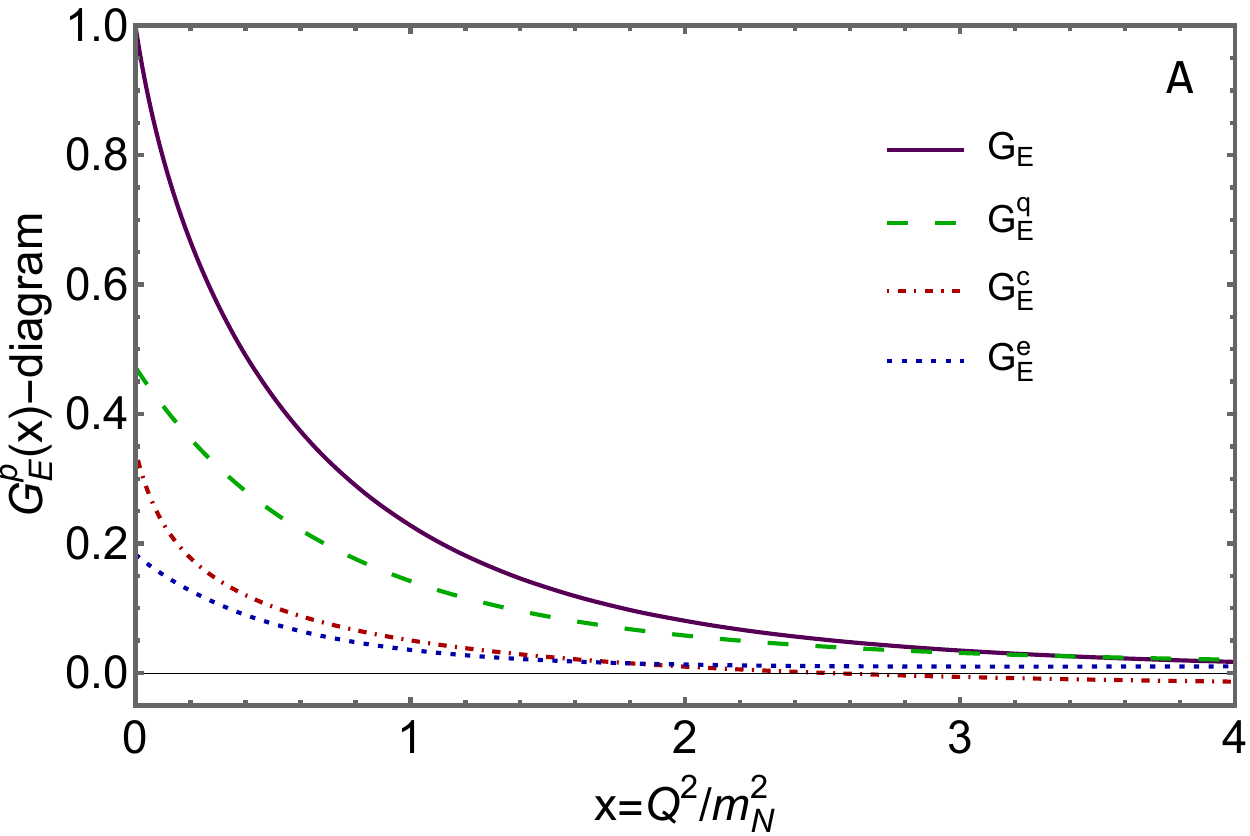}
\end{minipage}
\begin{minipage}[t]{0.48\textwidth}
\includegraphics[width=0.95\textwidth]{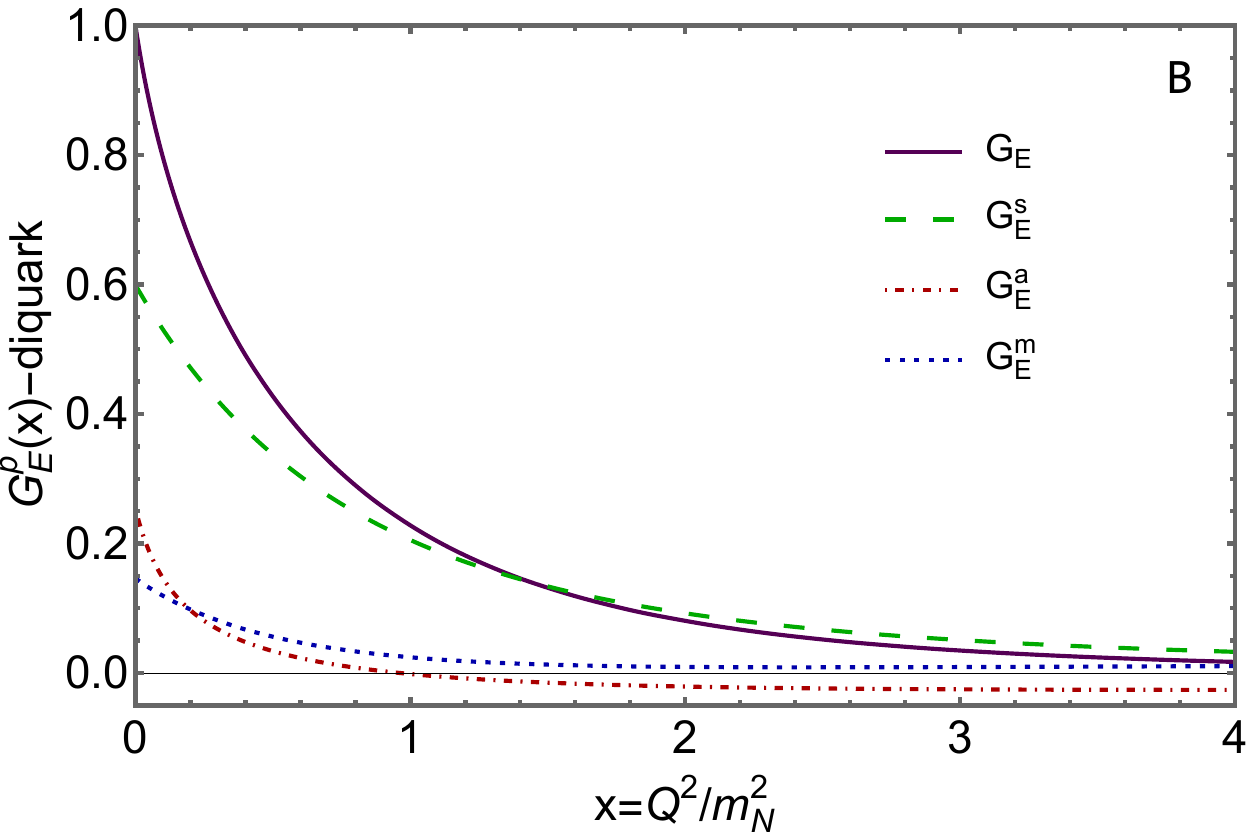}
\end{minipage}\vspace*{3ex}
\begin{minipage}[t]{0.48\textwidth}
\includegraphics[width=0.95\textwidth]{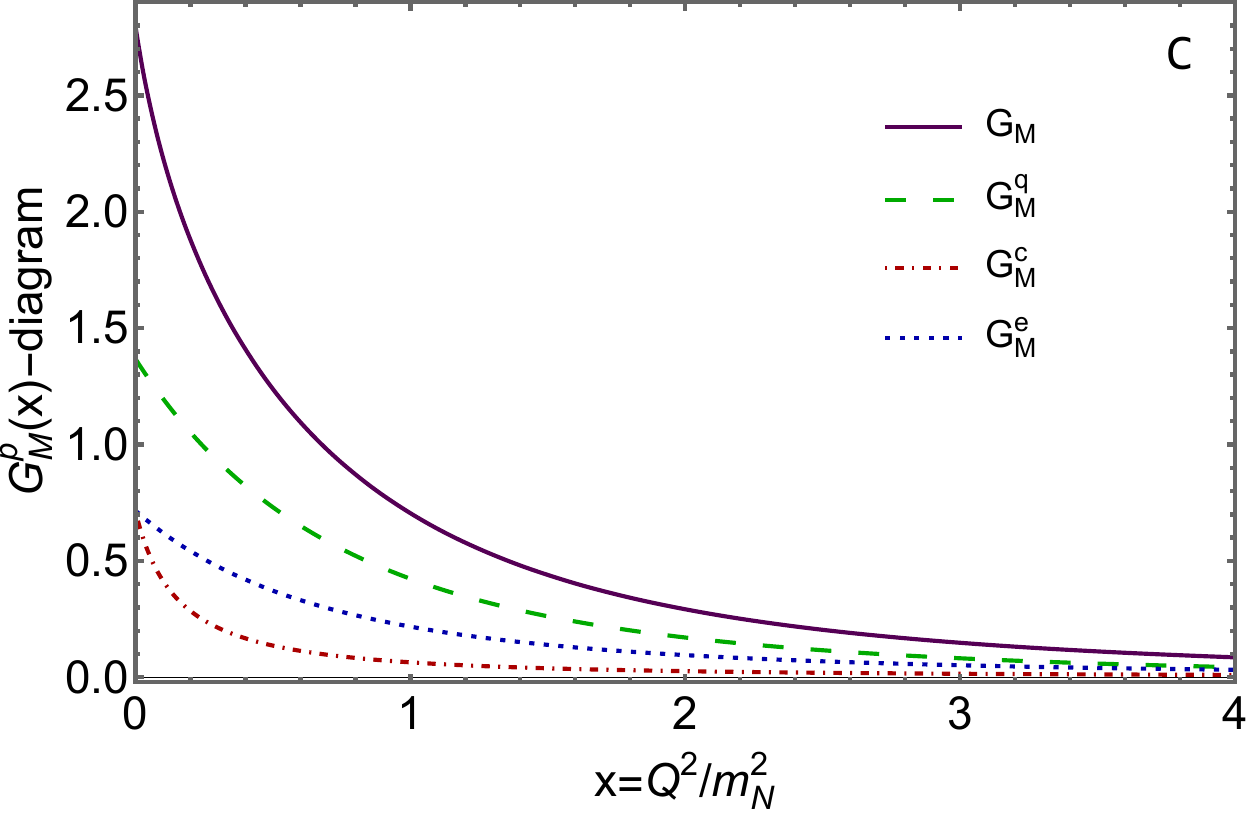}
\end{minipage}
\begin{minipage}[t]{0.48\textwidth}
\rightline{\includegraphics[width=0.95\textwidth]{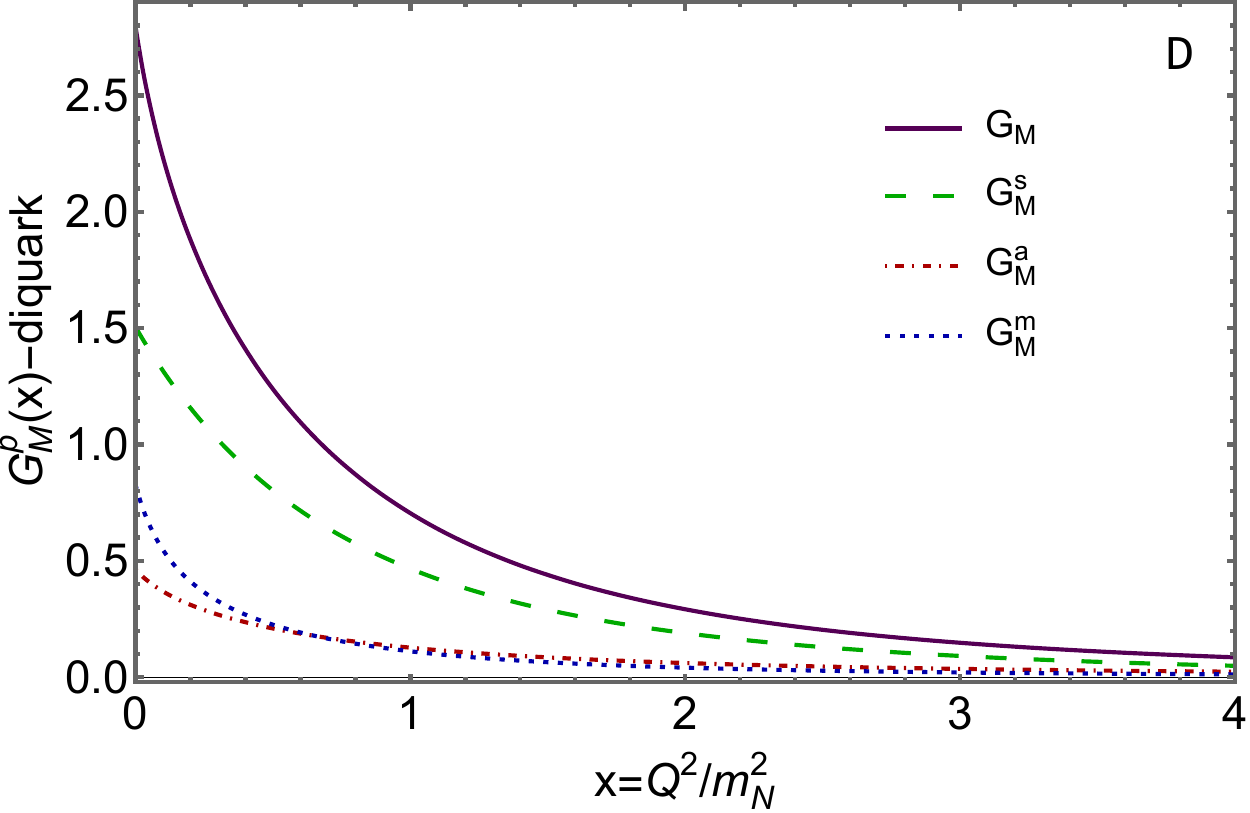}}
\end{minipage}
\end{minipage}
\caption{\label{FigProtonG}
Proton Sachs elastic form factors.
{\sf Panel A}. Electric, diagram breakdown.
{\sf Panel B}. Electric, diquark contributions.
{\sf Panel C}. Magnetic, diagram breakdown.
{\sf Panel D}. Magnetic, diquark contributions.
}
\end{figure*}

Our SPM extrapolations are developed as follows.\\[-4ex]
\begin{description}
\item[Step 1.]
For each function required, we produce $N=55$ directly calculated values, spaced evenly on $x \leq 6$.
\item[Step 2.]
From that set, $M_0=12$ points are chosen at random,
the usual SPM continued fraction interpolation is constructed,
and that function is extrapolated onto $6< x \leq 30$.
The curve is accepted so long as it is singularity free on $0\leq x \leq 30$.
\item[Step 3.]
Step~2 is repeated with another set of $M_0$ randomly chosen points.
With $M_0=12$, this procedure admits $\approx 5 \times 10^{10}$ independent interpolators.
\item[Step 4.]
One continues with 2 and 3 until $n_{M_0}=400\,$ smooth extrapolations are obtained.
\item[Step 5.]
Steps 2 and 3 are repeated for $M=\{M_0+ 2 i | i=1,\ldots,6\}$.
\item [Step 6.]
At this point, one has $3000$ statistically independent extrapolations of the subject form factor.
\end{description}
\vspace*{-1.0ex}
Working with these extrapolations, then at each value of $x$, we report the mean value of all curves as the central prediction and list as the uncertainty the function range which contains 68\% of all extrapolations: this is a $1\sigma$ band.

\section{Nucleon Electromagnetic Form Factors: Results}
\label{SecResults}
\subsection{Static properties}
Our results for nucleon static properties are reported in Table~\ref{StaticResults} alongside comparisons with the three-body predictions from Ref.\,\cite{Yao:2024uej} and empirical results taken from Ref.\,\cite[PDG]{ParticleDataGroup:2024cfk}.  Regarding the proton magnetic radius, although the PDG lists $r_M>r_E$, an independent SPM analysis \cite{Cui:2021skn} argues for $r_M<r_E$, with $r_M = 0.817(27)\,$fm.  An analogous analysis of the proton charge radius yields $r_E = 0.847(8)\,$fm \cite{Cui:2021vgm}.
Overall, compared with PDG values and allowing for the inflated value of $m_N$ herein, the mean absolute relative difference is $\overline{\rm ard} = 2.6(2.6)$.
Measured instead against the three-body results, $\overline{\rm ard} = 21.6(18.2)$.  The increase owes principally to the marked underestimate of nucleon magnetic moments in the three-body study.
Eliminating the nucleon magnetic moments, one finds $\overline{\rm ard} = 9.5(11.4)$.

\subsection{Form factors}
We will subsequently discuss the $x$-dependence of nucleon elastic electromagnetic form factors also presenting, in many cases, contribution separations according to the following legends.

\noindent Diagram contributions:\\[1ex]
\hspace*{1em}\parbox[b]{0.9\linewidth}{(\emph{q}) -- Diagram 1 -- contribution from the bystander quark;\\
(\emph{c}) -- Diagrams 2, 3 -- contribution from photon + diquark elastic or transition processes;\\
(\emph{e}) -- Diagrams 4\,-\,6 -- exchange/diquark breakup contributions.}

\noindent Diquark breakdown:\\[1ex]
\hspace*{1em}\parbox[b]{0.9\linewidth}{(\emph{s}) -- sum of all contributions that involve a scalar diquark in both $\Psi_{i,f}$;\\
(\emph{a}) -- sum of all contributions that involve an axialvector diquark in both $\Psi_{i,f}$;\\
(\emph{m}) -- sum of all contributions in which the diquark component of $\Psi_i$ differs from that in $\Psi_f$.}

\subsubsection{Proton Sachs}
Our results for proton Sachs form factors, Eq.\,\eqref{EqSachs}, are drawn in Fig.\,\ref{FigProtonG}.
Considering first the results on $x \simeq 0$ and working with Fig.\,\ref{FigProtonG}\,A, one may define an array of probabilities.
The probability that the interaction engages a bystander quark is $P_p^{q}=G_E^{q}(Q^2=0) = 47$\%;
that a diquark is struck, $P_p^{c} = G_E^{c}(Q^2=0) = 35$\%;
and that the photon interacts with the quark exchanged during diquark breakup and reformation,
$P_p^{e} = G_E^{e}(Q^2=0) = 18$\%.
Continuing in this vein, inspection of Fig.\,\ref{FigProtonG}\,B reveals the following probabilities:
photon interacts with scalar diquark component of the proton, $P_p^s = G_E^{s}(Q^2=0) = 60$\%;
photon interacts with axialvector diquark component, $P_p^a = G_E^{a}(Q^2=0) = 25$\%;
and photon flips the diquark content of the proton, $P_p^m = G_E^{m}(Q^2=0) = 15$\%.
This last set of probabilities indicates that the $0^+$ diquark only component constitutes 60\% of the proton's Poincar\'e covariant wave function.
Such a value delivers a ratio of valence-quark parton distribution functions that explains modern inferences from data \cite{JeffersonLabHallATritium:2021usd, Cui:2021gzg, Chang:2022jri}.
In fact, using Ref.\,\cite[Eq.\,(3)]{Roberts:2013mja}, the in-proton $d/u$ quark distribution function ratio at far-valence kinematics is $0.28$ \emph{cf}.\ $0.23(6)$ empirically \cite{JeffersonLabHallATritium:2021usd, Cui:2021gzg}.

Regarding the proton magnetic form factor, the photon + quark interaction produces 49\% of $\mu_p$;
photon + axialvector diquark scattering produces 25\%;
and photon-induced diquark breakup generates 26\%.
Turning to the diquark decomposition,
54\% of $\mu_p$ is associated with interactions that involve a scalar diquark in the initial and final state proton wave function;
16\% is associated with the analogous axialvector component;
and 30\% is generated by photon-induced flips of the proton's diquark content.

Some interesting observations attend upon the $x$ evolution of the form factors and their contributing components.
As will be discussed further below, $G_E^p$ possesses a zero.
Here we observe only that the driver for this outcome is photon + axialvector diquark elastic scattering.
This is evident in the following facts: $G_E^{pc}$ possess a zero at low $x$; $G_E^{pa}$ becomes negative at an even lower value; and both $G_E^{ps}$, $G_E^{pm}$ are positive definite.
It is worth noting that even though Refs.\,\cite{Cui:2020rmu, Segovia:2014aza, Cloet:2008re} also predict a zero in $G_E^p$ at a similar location, the cause is different.  In those studies, it is the appearance of a zero in $G_E^{pq}$ at low $x$ that drives the outcome; see Ref.\,\cite[Fig.\,7\,-\,top left]{Cloet:2008re}.

Regarding the proton magnetic form factor, the $G_M^{pc}$ term is positive definite and significant on $x\lesssim 0.5$.  As such, it is quantitatively distinct from the result in Refs.\,\cite{Cui:2020rmu, Segovia:2014aza, Cloet:2008re}, which is uniformly small in magnitude; see Ref.\,\cite[Fig.\,8\,-\,left]{Cloet:2008re}.  All other contributions are positive definite and significant on a larger domain.

Our proton Sachs form factors are compared with a compilation of available data in Fig.\,\ref{pGEGMdata}: the agreement is good.  There is a similar level of agreement with the $3$-body predictions delivered in Ref.\,\cite{Yao:2024uej}, which should be the case because these form factors were used in part to constrain the parameters in Table~\ref{list9parameters}.  Notably, the $3$-body predictions were made without reference to any form factor data.

\begin{figure}[t]
\includegraphics[clip, width=0.48\textwidth]{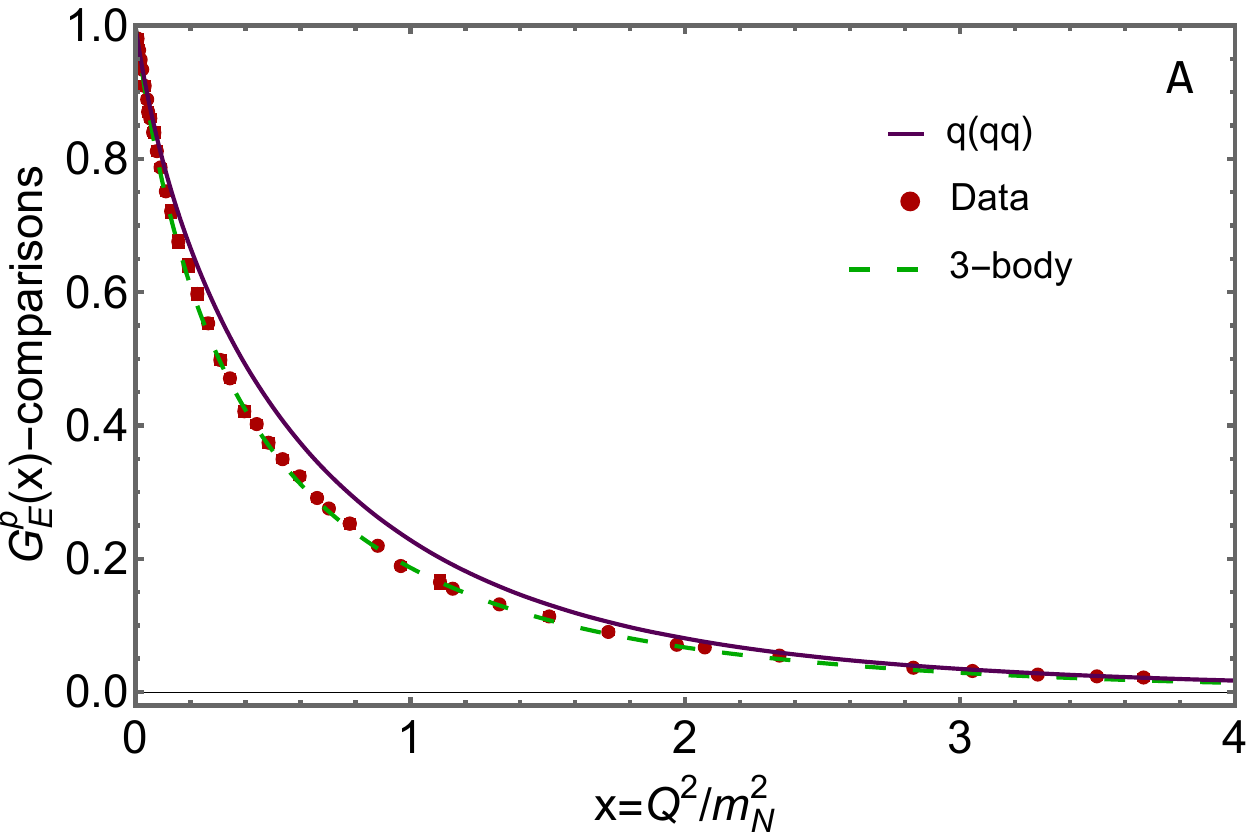}
\vspace*{1ex}

\includegraphics[clip, width=0.48\textwidth]{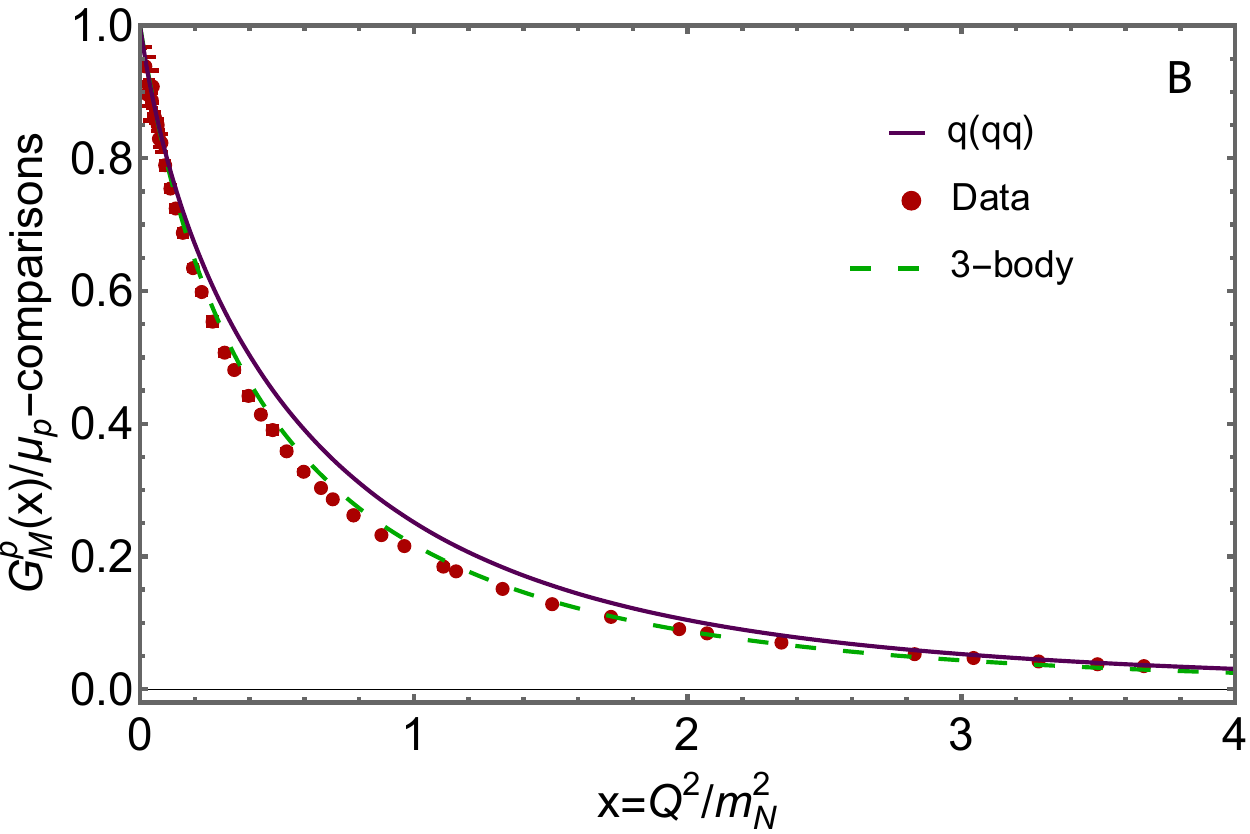}
\caption{\label{pGEGMdata}
Proton Sachs electric and magnetic form factors calculated herein, $q(qq)$, and compared with a compilation of available data \cite[Data]{Arrington:2007ux} (red circles) and predictions obtained using the $3$-body approach \cite[3-body]{Yao:2024uej} (dashed green curve).}
\end{figure}

The ratio $\mu_p G_E^p(x)/G_M(x)$ has been of great interest since the appearance of data that showed it to be a decreasing function of increasing $x$ \cite{Jones:1999rz}, especially after subsequent data confirmed the trend \cite{Gayou:2001qd, Punjabi:2005wq, Puckett:2010ac, Puckett:2017flj}.
Recently, an SPM analysis has shown that, independent of any model or theory of strong interactions and with a confidence level of $99.9$\%, these data are consistent with the existence of a zero in the ratio on $x<14.8$ \cite{Cheng:2024cxk}.

\begin{figure}[t]
\includegraphics[clip, width=0.48\textwidth]{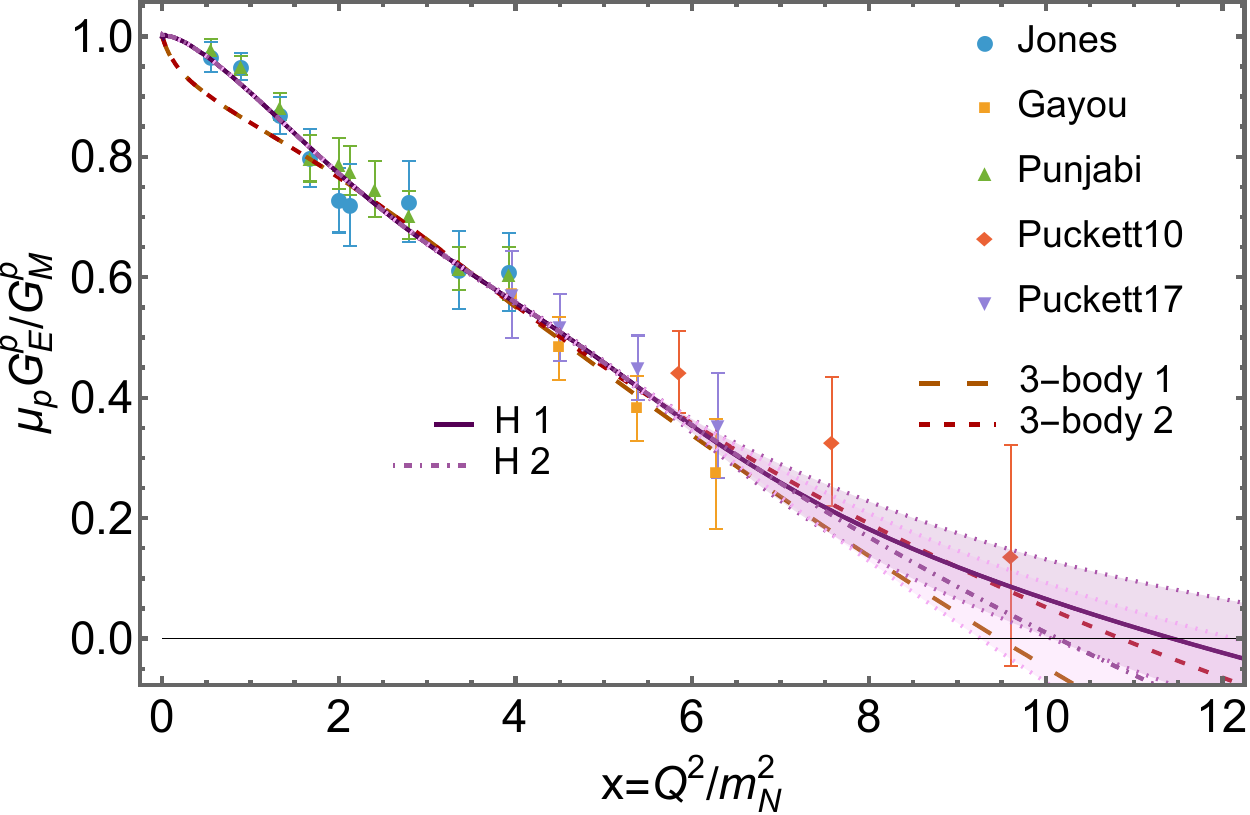}
\caption{\label{FigGEGM}
Result for proton ratio $\mu_p G_E^p/G_M^p$ calculated herein: H\,1 (solid purple) and H\,2 (dot-dashed violet), obtained using distinct SPM methods.  (The shaded bands indicate the $1\sigma$ SPM uncertainty; see Sect.\,\ref{Method}.)
Comparisons:
Data -- \cite[Jones]{Jones:1999rz}, \cite[Gayou]{Gayou:2001qd}, \cite[Punjabi]{Punjabi:2005wq}, \cite[ Puckett10]{Puckett:2010ac}, \cite[Puckett17]{Puckett:2017flj}; and
prediction obtained using the $3$-body approach \cite{Yao:2024uej} (long-dashed orange and dashed red distinguish the SPM methods).
}
\end{figure}

Our results for $\mu_p G_E^p(x)/G_M^p(x)$ are displayed in \linebreak Fig.\,\ref{FigGEGM}.  Since $r_E^p \gtrsim r_M^p$, this curve must fall slowly on $x\simeq 0$.
In order to deliver reliable results, free from numerical noise pollution, on $x\gtrsim 6$, we follow Ref.\,\cite{Yao:2024uej} and employ two distinct SPM extrapolation procedures.
Employing the SPM algorithm described in Sect.\,\ref{Method}, the schemes are as follows.
SPM\,1: Develop the ratio from SPM results for the individual numerator and denominator form factors.
SPM\,2: Develop the SPM extrapolation from the ratio $\mu_p G_E^p(x)/G_M^p(x)$ calculated on $x\leq 6$.
Referring to Fig.\,\ref{FigGEGM}, plainly, both analyses are mutually compatible.
Moreover, they agree with kindred analyses of predictions obtained using the three-body approach.

Importantly, all analyses predict a zero in $G_E^p$.
This outcome is meaningful because we constrained our electromagnetic vertex \emph{Ansatz} by this ratio only on $x \in (2.0,4.5)$.
A comparison of the zero locations follows:
\begin{equation}
\begin{array}{lll}
                  & q(qq)\;{\rm herein} & \rule{1ex}{0ex} {\rm 3-body} \\[1.5ex]
{\rm SPM\,1} \rule{1ex}{0ex} & 11.44_{-1.37}^{+3.35} & \rule{1ex}{0ex} 9.47_{-0.92}^{+1.90} \\[1.5ex]
{\rm SPM\,2} \rule{1ex}{0ex} & 10.14_{-0.88}^{+1.99} & \rule{1ex}{0ex} 10.85_{-0.96}^{+2.37}
\end{array}\,.
\end{equation}
All results are consistent with the practitioner-indepen\-dent inference from data that is described in Ref.\,\cite{Cheng:2024cxk}.  Combining our results, accounting for the asymmetric errors, the framework predicts a zero in $G_E^p$ at
\begin{equation}
x_z = 10.82_{-0.88}^{+1.60}\,.
\end{equation}
The 3-body result is $x_z = 10.46_{-0.70}^{+1.27}$.

\begin{figure*}[t]
\begin{minipage}[t]{\textwidth}
\begin{minipage}[t]{0.48\textwidth}
\includegraphics[width=0.95\textwidth]{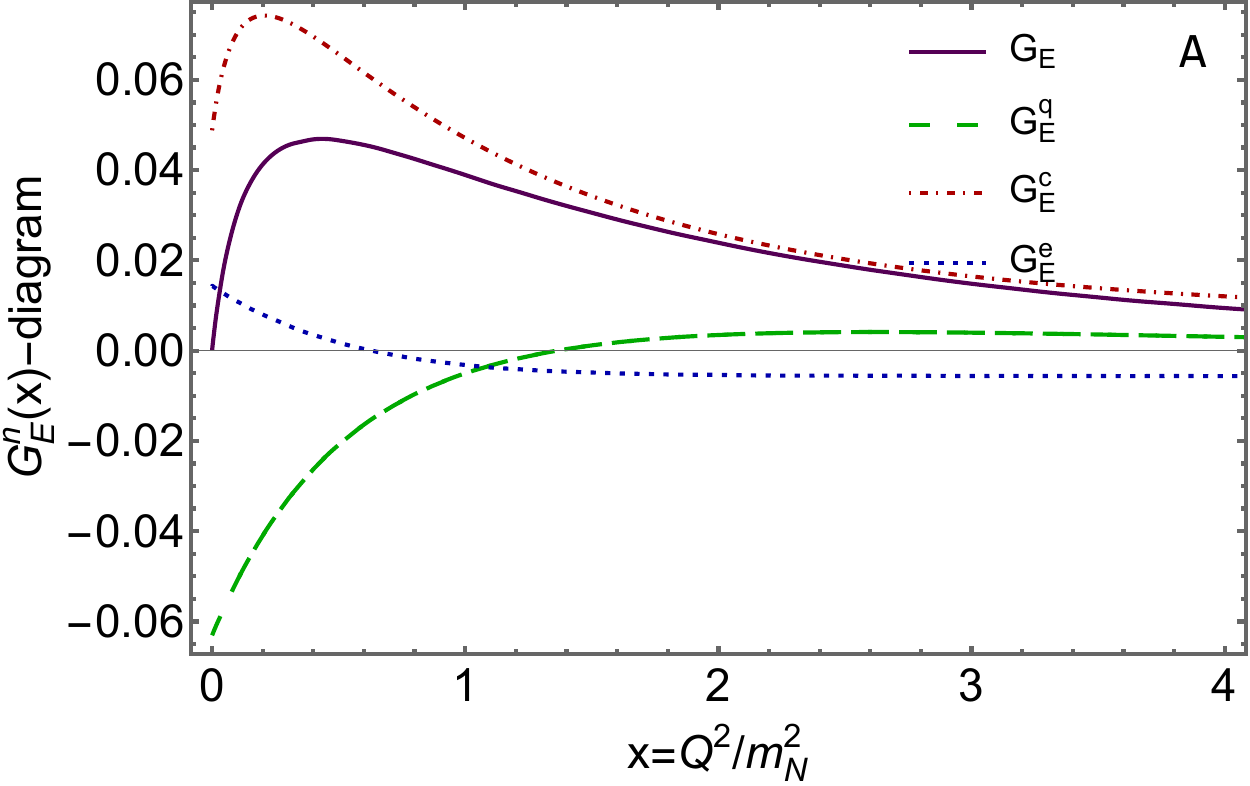}
\end{minipage}
\begin{minipage}[t]{0.48\textwidth}
\includegraphics[width=0.95\textwidth]{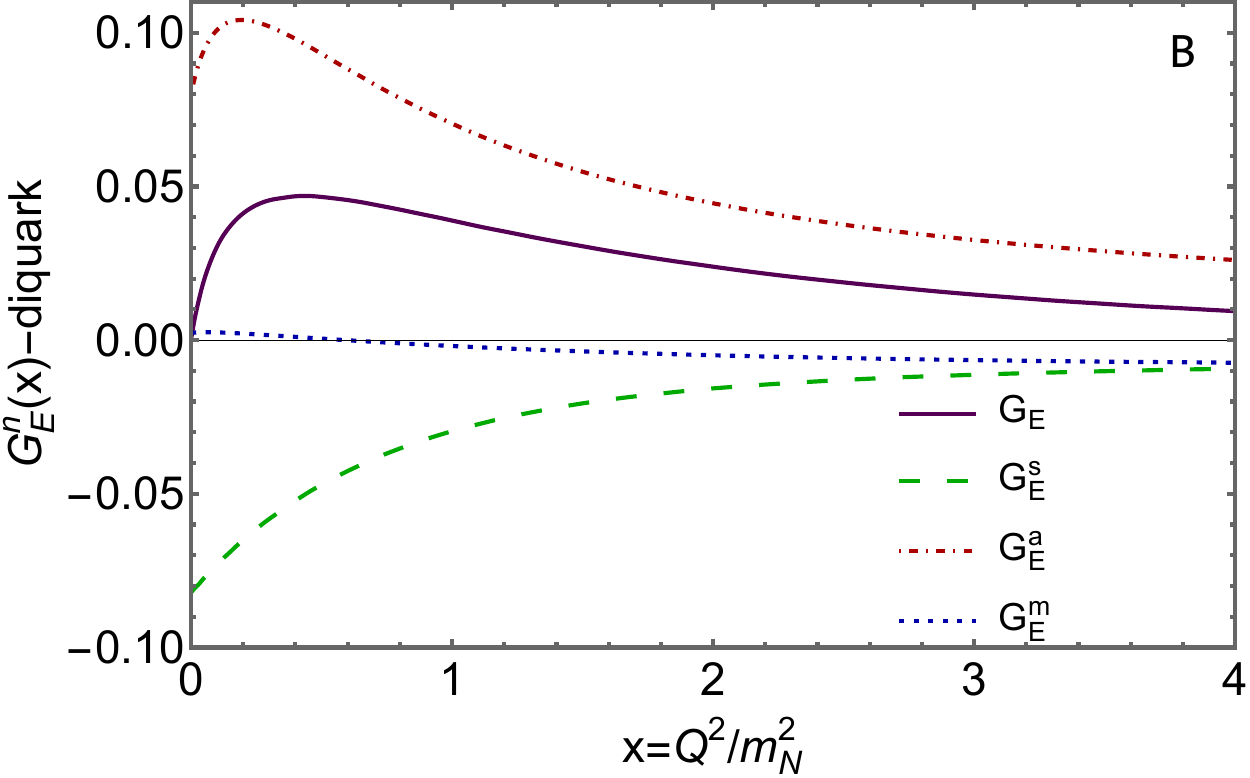}
\end{minipage}\vspace*{3ex}
\begin{minipage}[t]{0.48\textwidth}
\includegraphics[width=0.95\textwidth]{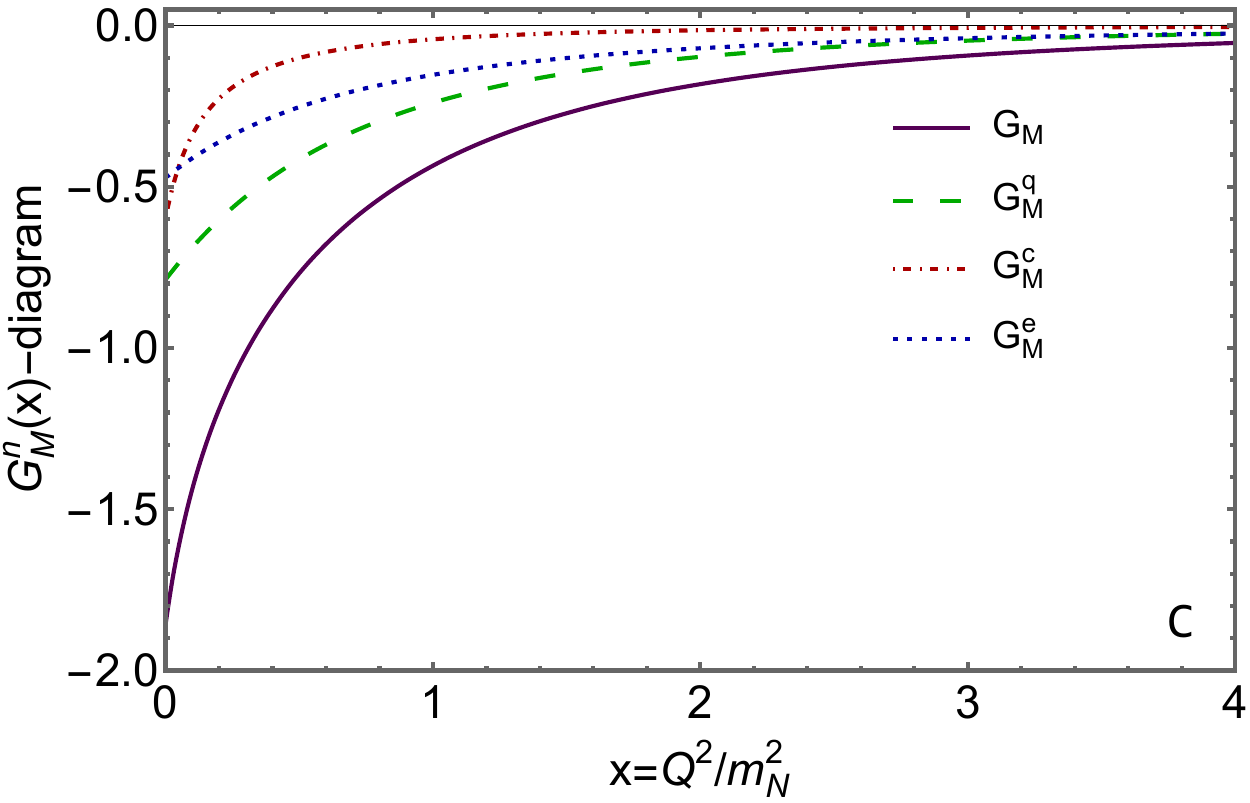}
\end{minipage}
\begin{minipage}[t]{0.48\textwidth}
\rightline{\includegraphics[width=0.95\textwidth]{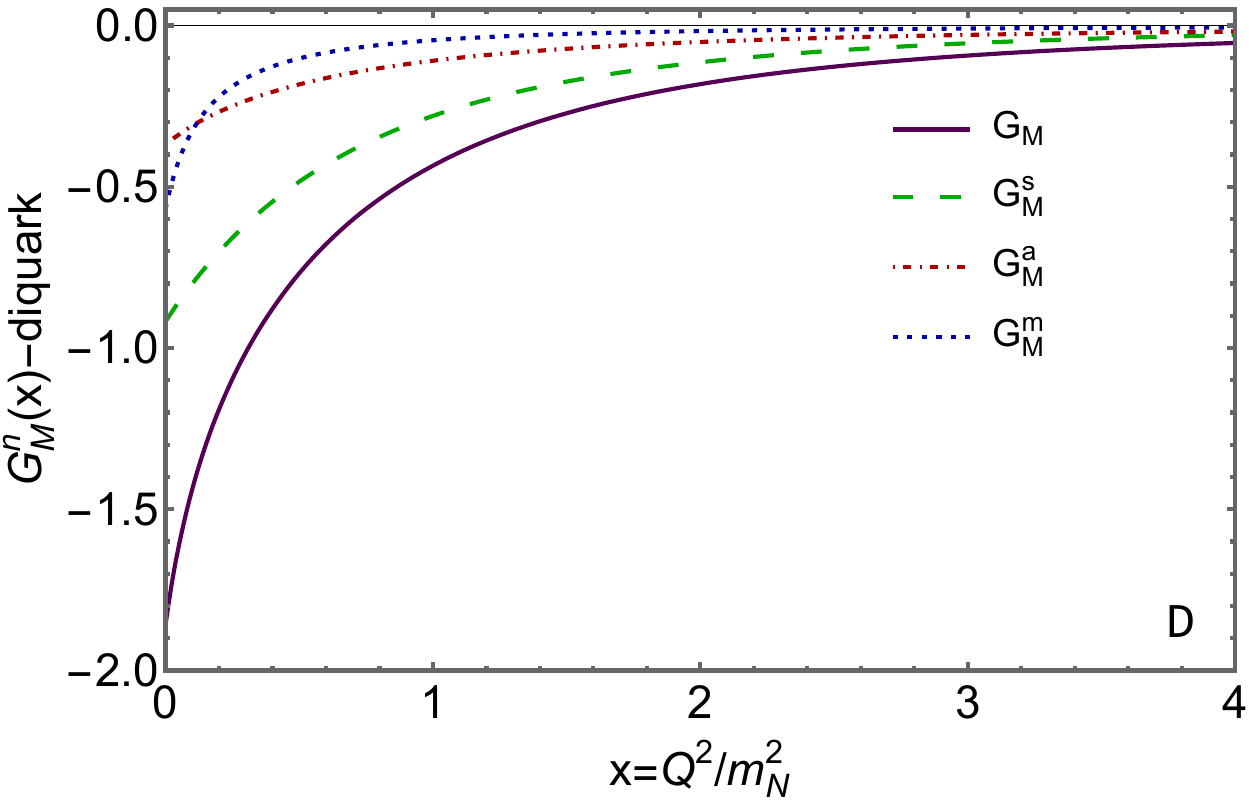}}
\end{minipage}
\end{minipage}
\caption{\label{FigNeutronG}
Neutron Sachs elastic form factors.
{\sf Panel A}. Electric, diagram breakdown.
{\sf Panel B}. Electric, diquark contributions.
{\sf Panel C}. Magnetic, diagram breakdown.
{\sf Panel D}. Magnetic, diquark contributions.
}
\end{figure*}

\subsubsection{Neutron Sachs}
Our results for neutron Sachs form factors, Eq.\,\eqref{EqSachs}, are drawn in Fig.\,\ref{FigNeutronG}.
Considering first the results on $x \simeq 0$ and working with Fig.\,\ref{FigNeutronG}\,A, it is plain that current conservation is implemented perfectly: the neutron is neutral.
Moreover, the active quark contribution is relatively large in magnitude and negative because, at long range, this is a negatively charged $d$ quark associated with a bystander $[ud]$ diquark.
As $x$ increases, \emph{i.e}., one probes more toward the neutron interior, the interaction can also involve a positively charged $u$ quark in the presence of a $\{dd\}$ bystander diquark; so, the contribution changes sign.
The correlation contribution is positive definite largely because $[ud]$, $\{ud\}$ diquarks are positively charged and $[ud]$ is the dominant correlation.
Informed by Fig.\,\ref{FigNeutronG}\,B, which indicates that $G_E^n$ is dominated by contributions with the same diquark in the initial and final state, then the exchange/breakup contribution is positive at long range (small $x$) because $d[ud] \to [du] d$ involves the exchange of a positively charged $u$ quark.
It changes sign as $x$ increases (the length scale diminishes) because the heavier axialvector diquark comes to dominate, $\{dd\}$ is more likely than $\{ud\}$, so exchange of the negative $d$ quark is more probable.

Turning to Fig.\,\ref{FigNeutronG}\,B, the $G_E^n$ contribution from scalar diquark in both initial and final state neutron is negative definite because the hardest scatterer is the active, negatively charged $d$ quark.
$G_E^a$ is positive definite because, owing to current conservation, scattering from the $u$ in the $u\{dd\}$ configuration must dominate on $x\simeq 0$; and since the $u\{dd\}$ configuration is always more heavily weighted than $d\{ud\}$, the bystander $u$ quark remains the dominant hard scatterer as $x$ increases.

Regarding the neutron magnetic form factor, the photon + quark interaction produces 42\% of $\mu_n$;
photon + axialvector diquark scattering produces 32\%;
and photon-induced diquark breakup generates 26\%.
Turning to the diquark decomposition,
49\% of $\mu_n$ is associated with interactions that involve a scalar diquark in the initial and final state proton wave function;
20\% is associated with the analogous axialvector component;
and 31\% is generated by photon-induced flips of the proton's diquark content.
These results indicate that both the neutron and proton anomalous magnetic moments are very similar in origin.

\begin{figure}[t]
\includegraphics[clip, width=0.48\textwidth]{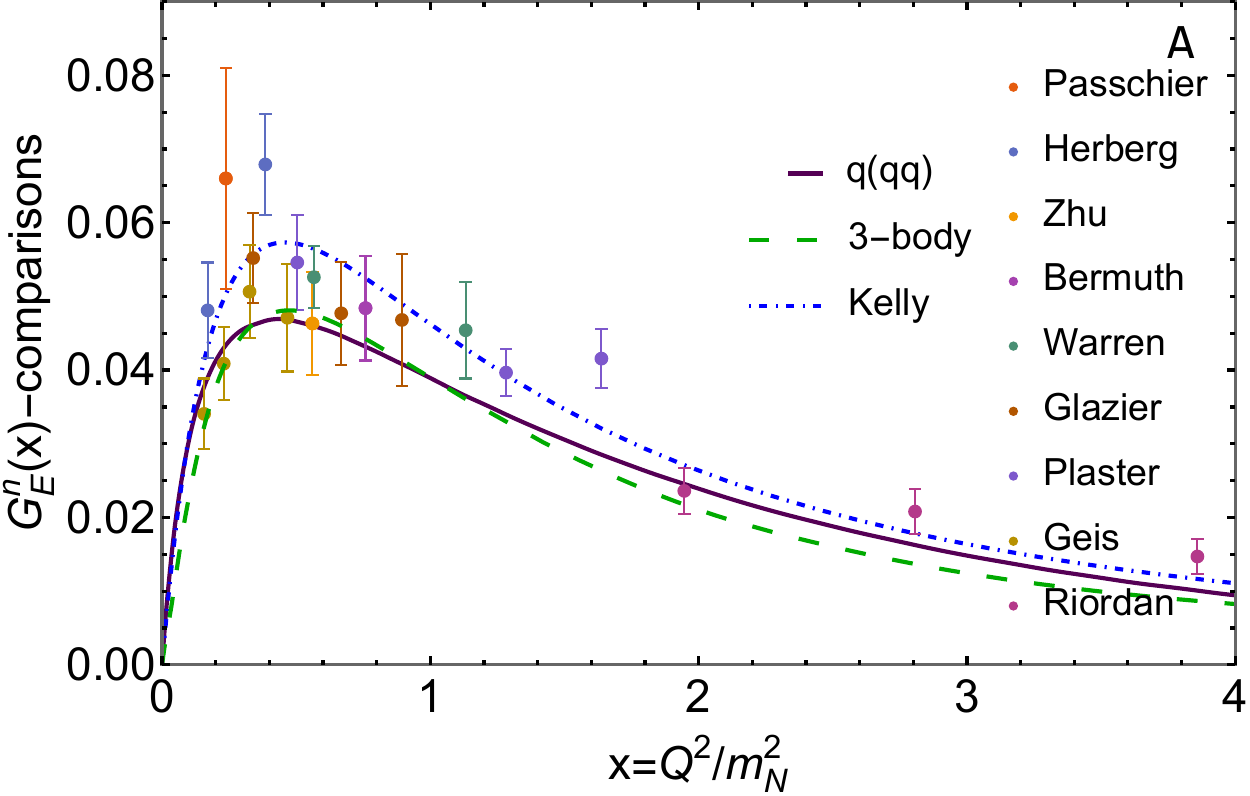}
\vspace*{1ex}

\includegraphics[clip, width=0.48\textwidth]{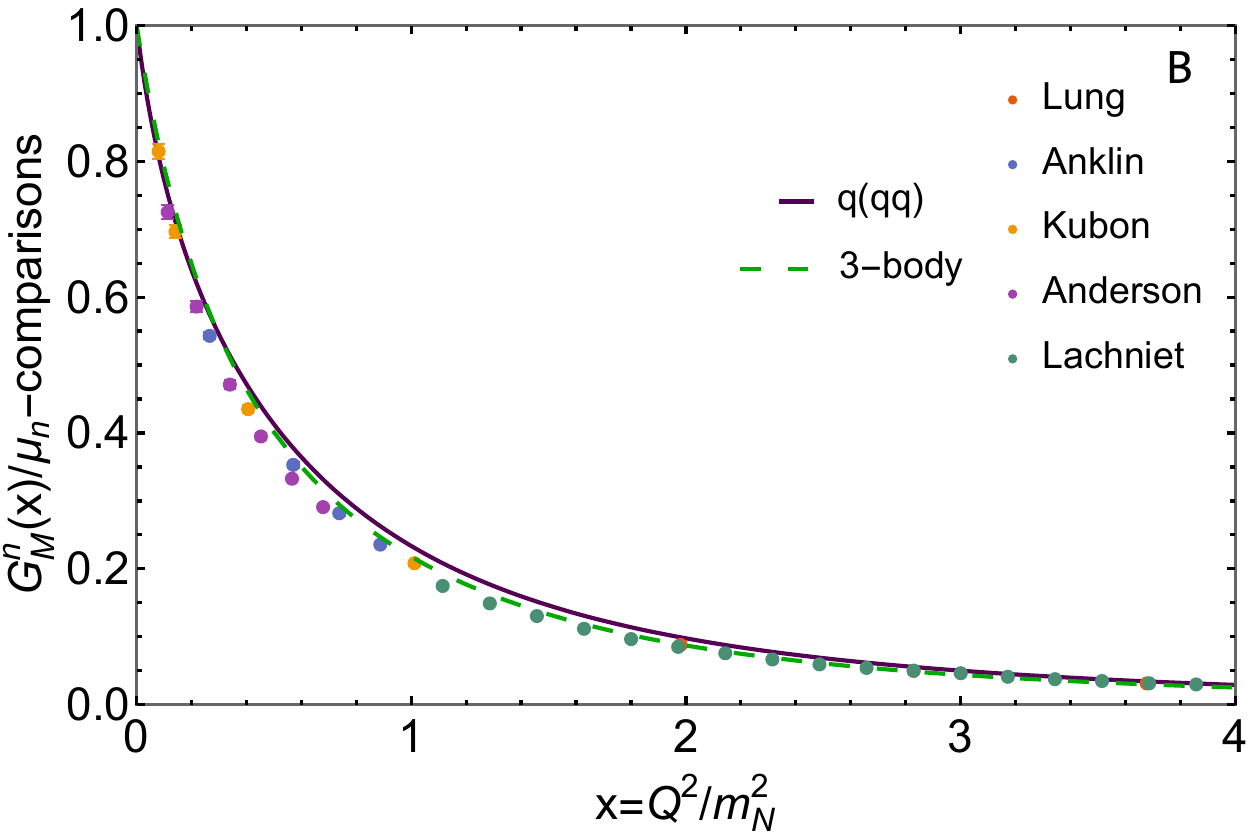}
\caption{\label{FigNeutronSachs}
{\sf Panel A}. Neutron Sachs electric form factor calculated herein ($q(qq)$, solid purple curve).
It is compared with the prediction obtained using the $3$-body approach \cite[3-body]{Yao:2024uej} (long-dashed green) and the parametrisation of data in Ref.\,\cite[Kelly]{Kelly:2004hm} (dot-dashed blue).
The data depicted data are from the following sources:
Refs.\,\cite{Passchier:1999cj, Herberg:1999ud, E93026:2001css, Bermuth:2003qh, Warren:2003ma, Glazier:2004ny, Plaster:2005cx, BLAST:2008bub, Riordan:2010id}.
{\sf Panel B}.
Neutron Sachs magnetic form factor calculated herein  (solid purple curve)
compared with the prediction obtained using the $3$-body approach \cite{Yao:2024uej} (dashed green curve).
The data depicted data are from the following sources:
Refs.\,\cite{Lung:1992bu, Anklin:1998ae, Kubon:2001rj, JeffersonLabE95-001:2006dax, CLAS:2008idi}.
}
\end{figure}

Our predictions for the neutron Sachs form factors are drawn in Fig.\,\ref{FigNeutronSachs} and compared therein with predictions from the $3$-body approach \cite{Yao:2024uej} and available data \cite{Passchier:1999cj, Herberg:1999ud, E93026:2001css, Bermuth:2003qh, Warren:2003ma, Glazier:2004ny, Plaster:2005cx, BLAST:2008bub, Riordan:2010id, Lung:1992bu, Anklin:1998ae, Kubon:2001rj, JeffersonLabE95-001:2006dax, CLAS:2008idi}.
Compared with earlier quark + diquark analyses \cite{Cui:2020rmu, Segovia:2014aza, Cloet:2008re}, our description of $G_E^n$ is significantly better; and it is practically equivalent to that produced by the recent $3$-body study \cite{Yao:2024uej}.  The relative ${\mathpzc L_1}$-difference between our prediction and the data fit in Ref.\,\cite{Kelly:2004hm} is 14\%.
Our prediction for $G_M^n$ matches well with data.

\begin{figure}[t]
\includegraphics[clip, width=0.48\textwidth]{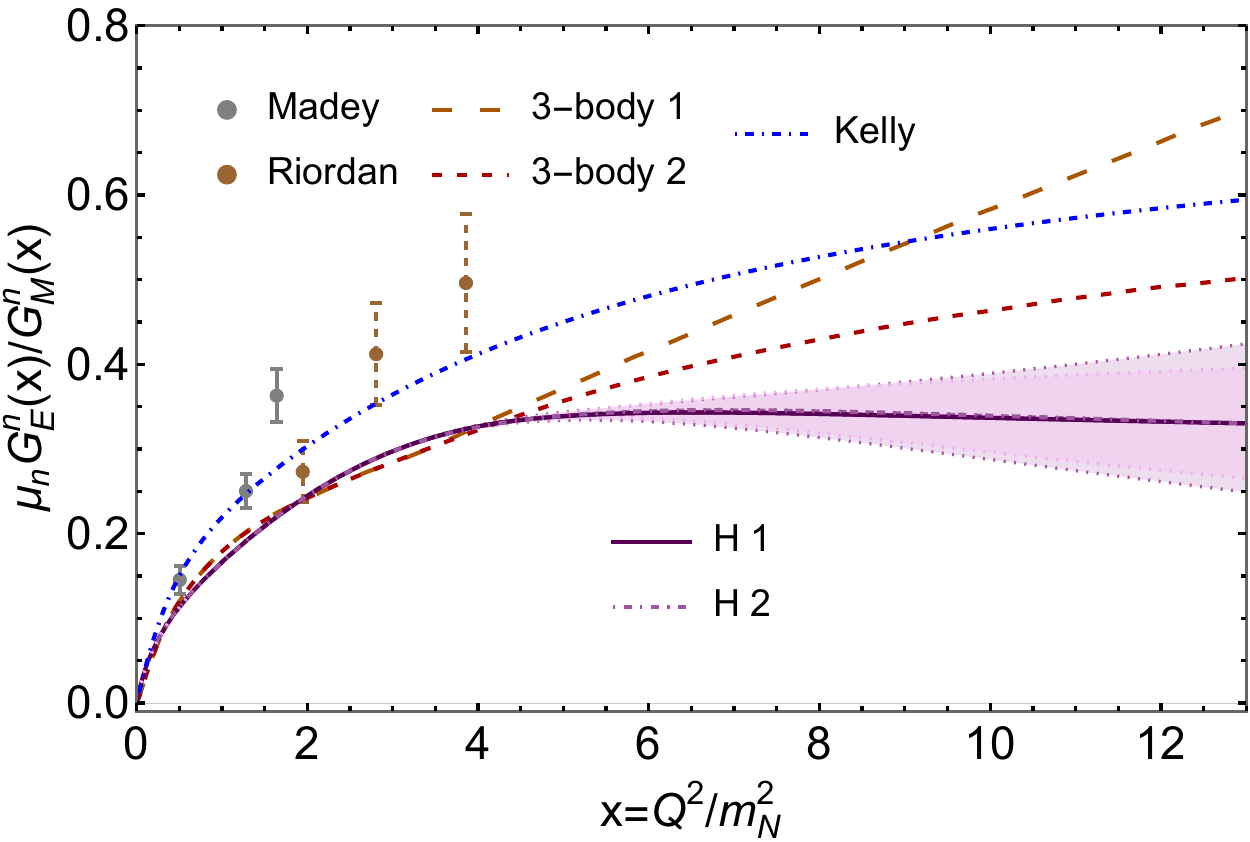}
\caption{\label{FigGEGMn}
Result for neutron ratio $\mu_n G_E^n/G_M^n$ calculated herein: H\,1 (solid purple) and H\,2 (dot-dashed violet), obtained using distinct SPM methods.
Comparisons:
predictions obtained using the $3$-body approach \cite[3-body]{Yao:2024uej} (long-dashed orange and dashed red distinguish the SPM methods); and parametrisation of data in Ref.\,\cite[Kelly]{Kelly:2004hm} (dot-dashed blue).
Data:  \cite[Riordan]{Riordan:2010id}, \cite[Madey]{Madey:2003av}.
}
\end{figure}

We display our prediction for $\mu_n G_E^n(x)/G_M^n(x)$ in Fig.\,\ref{FigGEGMn}.
As with the analogous proton ratio, in order to deliver a curve free from numerical noise, on $x\gtrsim 6$ we employ the same two distinct SPM extrapolation procedures.  Plainly, they deliver mutually consistent results.
The agreement with data is fair.
Regarding comparison with the $3$-body analyses \cite{Yao:2024uej}, deviations become evident on $x\gtrsim 5$.
Evidently and unsurprisingly, the mismatches owe largely to differences in $G_E^n$.
On the other hand, our predictions and those in Ref.\,\cite{Yao:2024uej} agree in finding no zero crossing in $G_E^n$ on $x\lesssim 15$.
In this they are consistent with Ref.\,\cite{Cui:2020rmu}, but disagree with the earlier $q(qq)$ studies in Ref.\,\cite{Segovia:2014aza, Cloet:2008re}, in which the numerical algorithms were less refined, making no use of the SPM to reach large $x$.

\begin{figure*}[t]
\begin{minipage}[t]{\textwidth}
\begin{minipage}[t]{0.48\textwidth}
\includegraphics[width=0.95\textwidth]{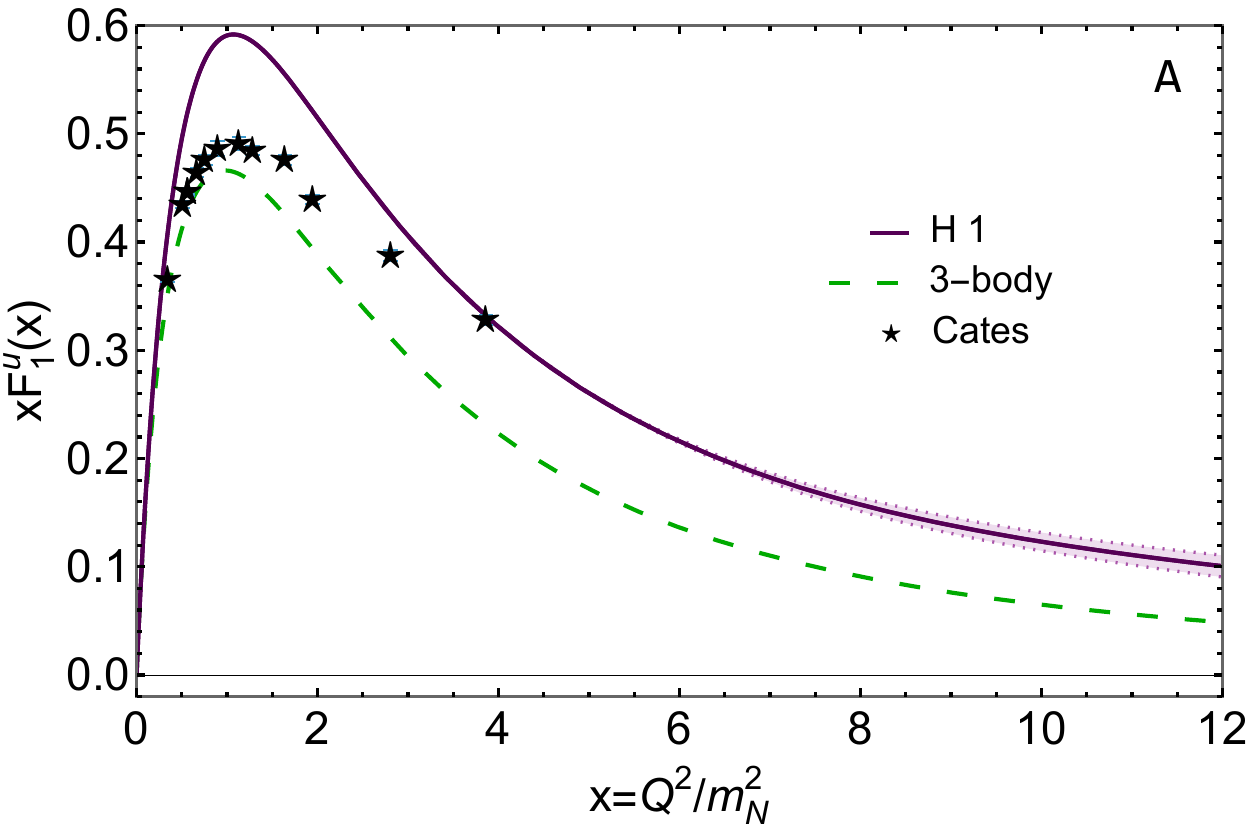}
\end{minipage}
\begin{minipage}[t]{0.48\textwidth}
\includegraphics[width=0.95\textwidth]{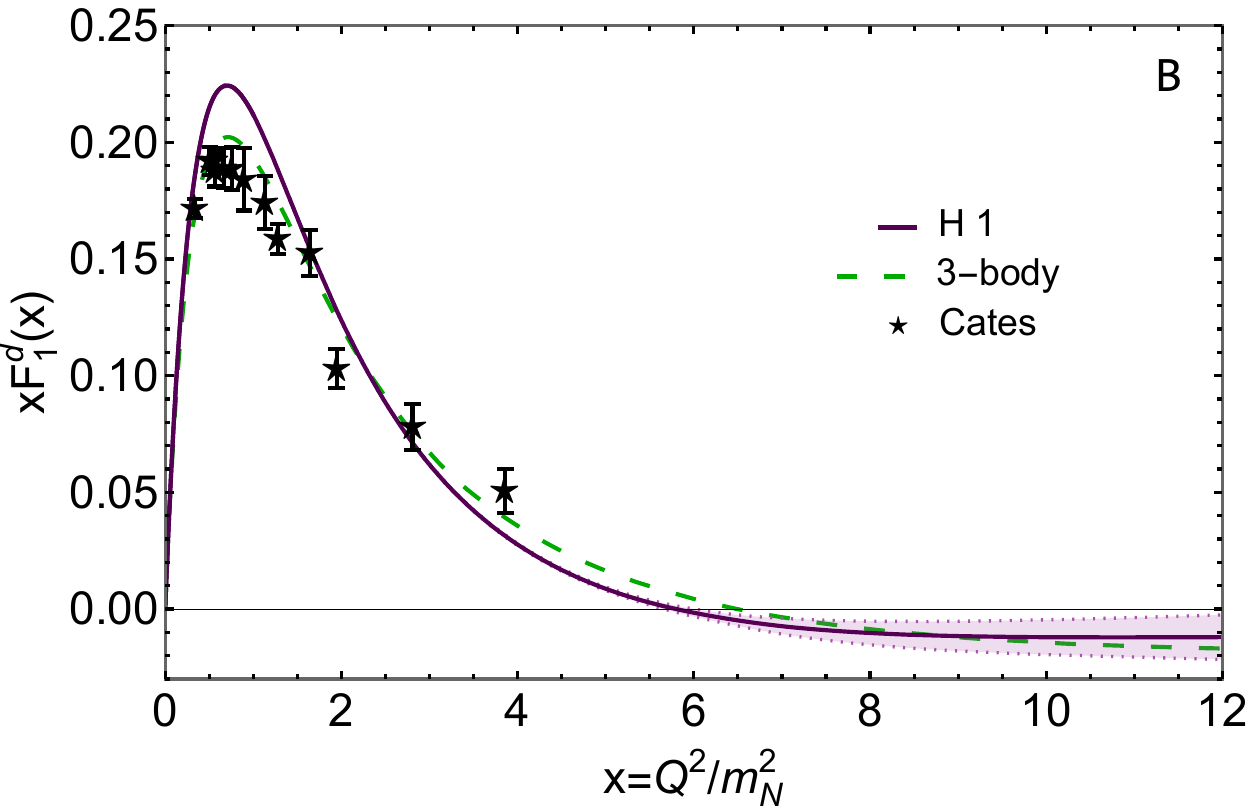}
\end{minipage}\vspace*{3ex}
\begin{minipage}[t]{0.48\textwidth}
\includegraphics[width=0.95\textwidth]{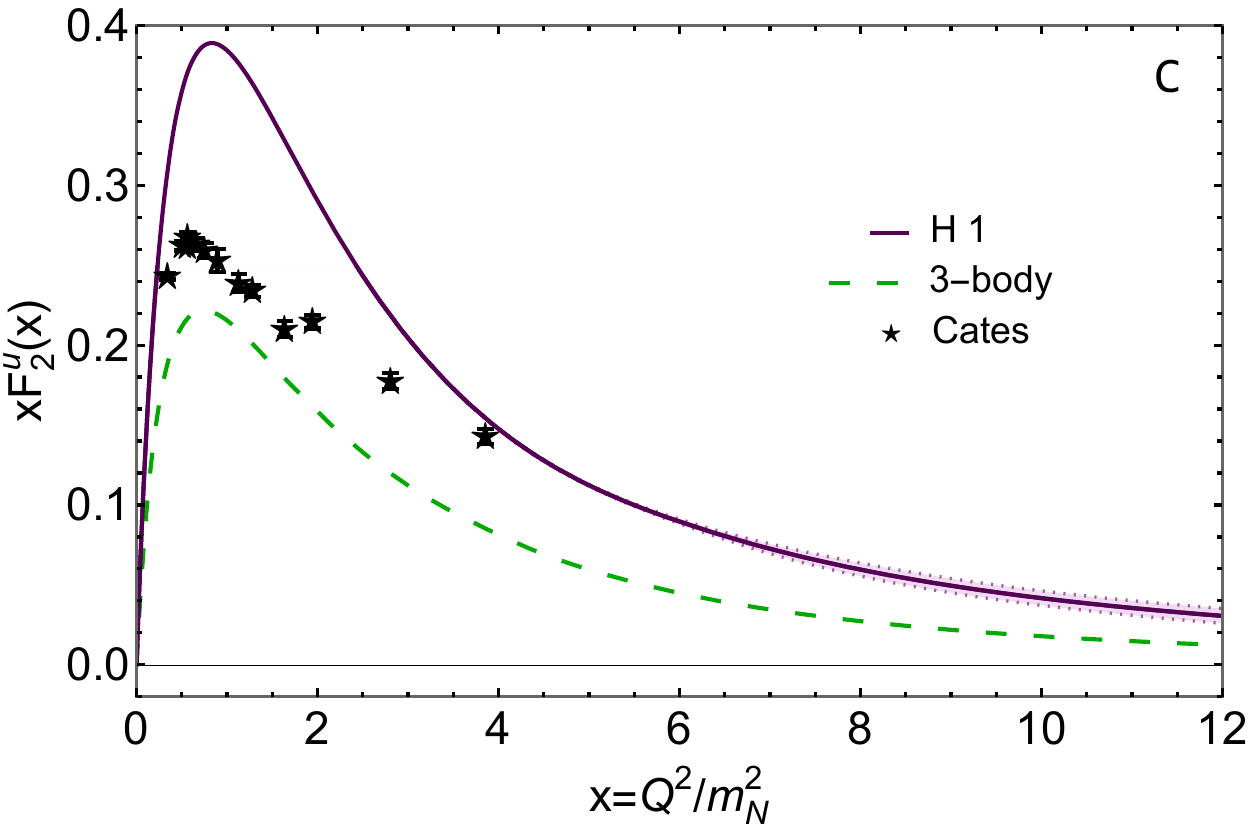}
\end{minipage}
\begin{minipage}[t]{0.48\textwidth}
\rightline{\includegraphics[width=0.95\textwidth]{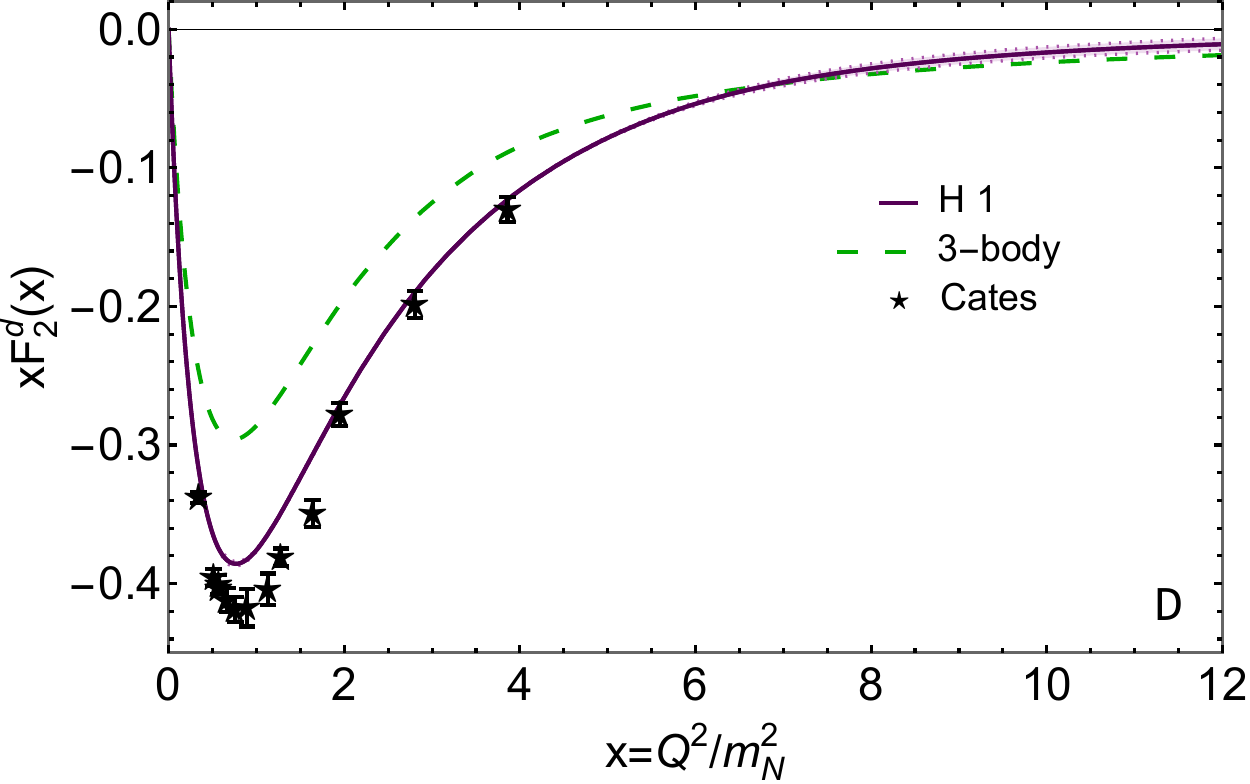}}
\end{minipage}
\end{minipage}
\caption{\label{FSep}
Flavour separated form factors -- solid purple curve and like-coloured SPM uncertainty band, H\,1.
{\sf Panel A}. Dirac, $u$ quark.
{\sf Panel B}. Dirac, $d$ quark.
{\sf Panel C}. Pauli, $u$ quark.
{\sf Panel D}. Pauli, $d$ quark.
Comparisons:
$3$-body results from Ref.\,\cite[[3-body]{Yao:2024uej} -- long dashed green curve;
data -- Ref.\,\cite[Cates]{Cates:2011pz}.
}
\end{figure*}

\section{Nucleon Electromagnetic Form Factors: Flavour Separation}
\label{SecFlavour}
\subsection{Dirac and Pauli}
Precise measurements of $G_E^n$ to $x =3.9\,$ \cite{Riordan:2010id} enabled a flavour separation of the nucleon form factors on a significant momentum domain \cite{Cates:2011pz}.  Assuming one can neglect $s$-quark contributions, which is true by definition at $\zeta_{\cal H}$ and a good approximation at higher resolving scales, then the Dirac and Pauli form factors can be decomposed as follows:
\begin{equation}
\label{FlavourSep}
F_{i}^u = 2 F_{i}^p + F_{i}^{n}, \;
F_{i}^d = F_{i}^{p} + 2 F_{i}^{n} \,, i=1,2.
\end{equation}
Evidently,
\begin{equation}
\label{FlavourNorm}
F_{1}^{u}(x=0)=2\,, \; F_{1}^{d}(x=0)=1\,.
\end{equation}

Our predictions for these flavour separated form factors are displayed in Fig.\,\ref{FSep}.  They reproduce the trends seen in extant data, agree in sign and also fairly well in magnitudes.  Importantly, the analysis predicts that $F_{1}^d$ exhibits a zero:
\begin{equation}
\label{F1dzero}
F_1^d(x_{z_1}) = 0\,, x_{z_1} = 5.80_{-0.14}^{+0.20}\,.
\end{equation}
Such a zero was also seen in earlier quark + diquark analyses \cite{Segovia:2014aza, Cui:2020rmu}, although it was located at a 40\% larger value of $x$.
The $3$-body study finds a zero in $F_1^d$ at $x \approx 6.5$.
Alike with that study, we find that no other flavour separated form factor possesses a zero.
In particular, the zero found by Ref.\,\cite{Cui:2020rmu} in $F_2^d$ is not reproduced.
Results from analyses of new data are soon expected to test Eq.\,\eqref{F1dzero}.

Reviewing the upper panels of Fig.\,\ref{FSep}, one sees that, even allowing for the difference in normalisation, \linebreak Eq.\,\eqref{FlavourNorm}, $F_1^d$ is smaller than $F_1^u$ and decreases more quickly as $x$ increases; obviously, because it possesses a zero.
This behaviour can be understood by noting that the proton's Faddeev wave function is dominated by the $[ud]$ scalar diquark correlation, which produces $\approx 60$\% of the proton's normalisation -- Fig.\,\ref{FigProtonG}\,B.  Hence, $ep$ scattering is dominated by the virtual photon striking the $u$-quark that is not participating in the correlation.  Scattering from the valence $d$-quark is suppressed because this quark is usually absorbed into a soft correlation.  The effect is most noticeable at large $x$.

In order to explain the zero in $F_1^d$, recall that an active axialvector diquark component is also present in the proton.
It occurs in two combinations $u\{ud\}$ and $d\{uu\}$.
The second ensures that, although with less overall likelihood, valence $d$-quarks are always available to participate in a hard scattering event; consequently, this piece provides the leading contribution to $F_1^d$ on $x\gtrsim 2$.
With increasing $x$, $F_1^d$ passes through zero because of interference between the various diquark-component contributions to $e d$-quark scattering within the proton.
So, like the ratio of quark structure functions on the valence domain, the location of the zero in $F_1^d$ is sensitive to the relative probability of finding axialvector and scalar diquarks in the proton.
Analogous effects are likely generated dynamically in the $3$-body analysis, but a detailed exposition is not yet available.

Notably, the existence of a zero in $F_1^d$ highlights that any empirical suggestion of power-law scaling in nucleon electromagnetic form factors on the currently accessible $x$ domain is likely incidental because
the zero expresses a continuing role for correlations that distinguish between quark flavours and impose different features upon their scattering patterns.

The lower panels of Fig.\,\ref{FSep} depict the proton's flavour-separated Pauli form factors.
The remarks made in connection with the behaviour of $F_1^{u,d}$ are also largely applicable here and their explanations are similar.
However, there is one significant difference, \emph{viz}.\ despite the fact that the $d/u$ ratio in the proton is $2/1$, the $u$- and $d$-quark Pauli form factors are roughly equal in magnitude on $x\lesssim 5$.  We return to this in Sect.\,\ref{SecAM}.

\subsection{Light-front-transverse densities}
\label{SubSecLFTDs}
\subsubsection{Number}
The flavour separated elastic form factors provide a measure of the $Q^2$-dependence of valence $u$- and $d$-quark elastic scattering probabilities within the proton.
Exploiting the properties of generalised parton distributions in impact-parameter space, the following two-dimensional Fourier transforms express light-front-tran\-sverse valence-quark densities \cite{Burkardt:2002hr, Diehl:2003ny, Mezrag:2023nkp} $(f=u,d)$:
\begin{equation}
\label{fdensity}
 \hat\rho_1^f(|\hat b|) = \int\frac{d^2 \vec{q}_\perp}{(2\pi)^2}\,{\rm e}^{i \vec{q}_\perp \cdot \hat b}
F_1^{f}(Q^2)\,,
\end{equation}
with $F_1^f(Q^2)$ interpreted in the frame defined by $Q^2 = m_N^2 |q_\perp|^2$, $m_N q_\perp =  (Q_1,Q_2,0,0)$.  Such densities have also been considered elsewhere; see, \emph{e.g}., Refs.\,\cite{Miller:2010nz, Mondal:2015uha, Cui:2020rmu}.

\begin{figure}[t]
\includegraphics[clip, width=0.48\textwidth]{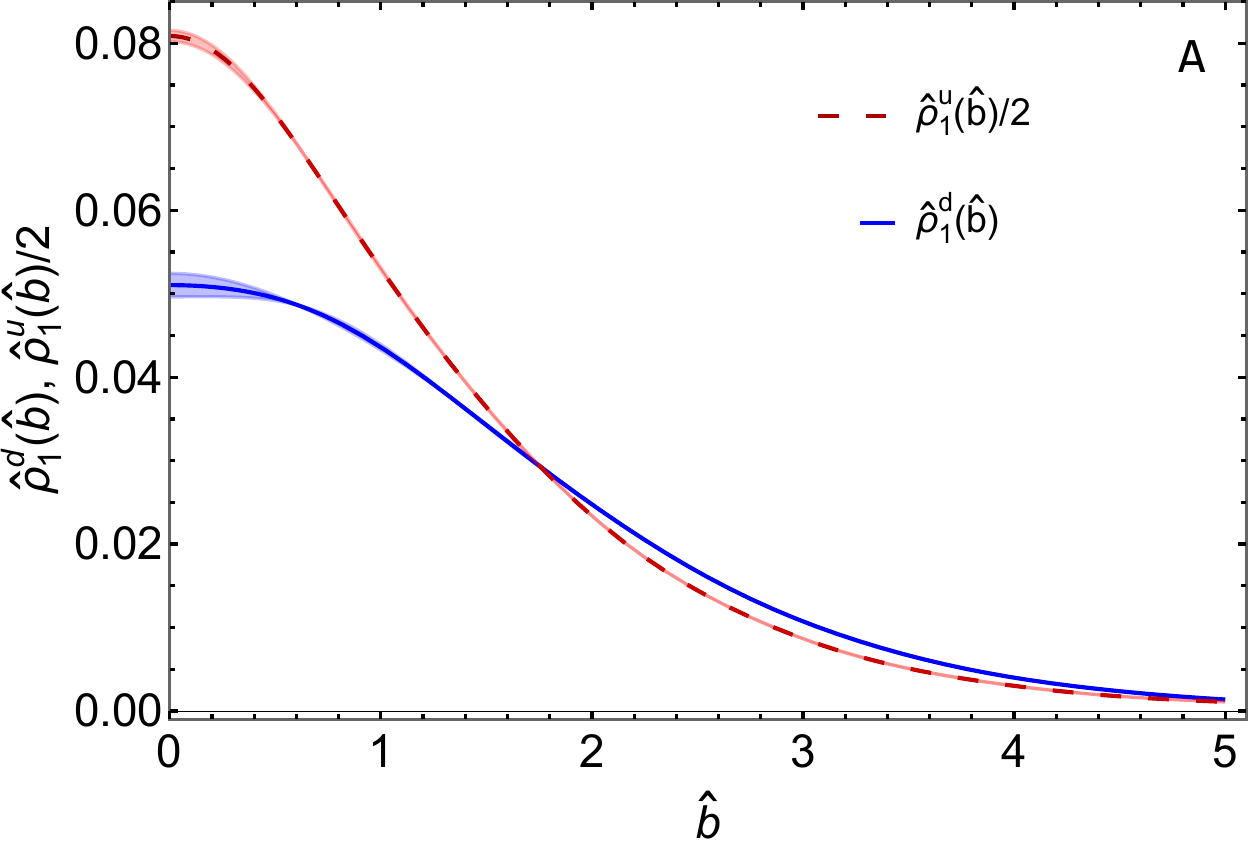}
\vspace*{1ex}

\includegraphics[clip, width=0.48\textwidth]{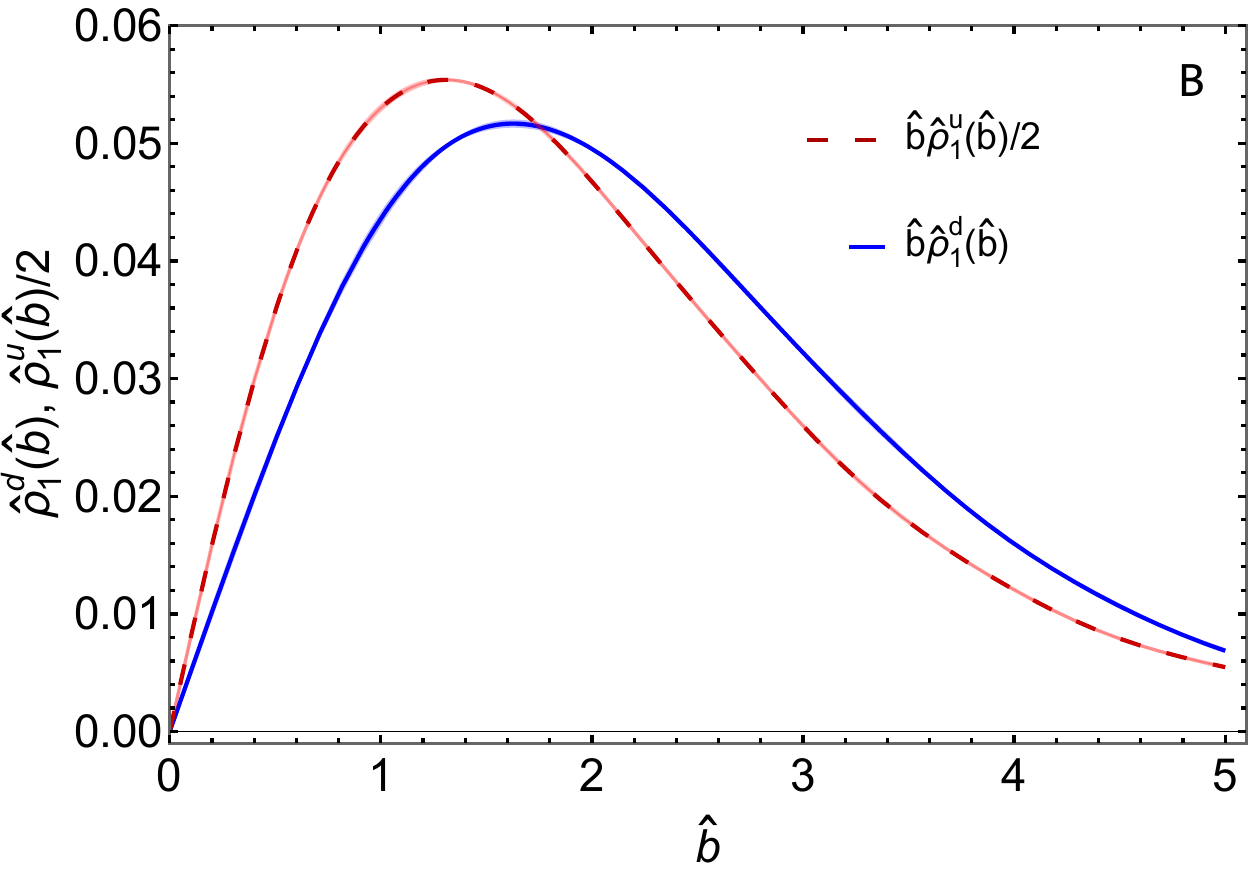}
\caption{\label{Figfdensity}
Light-front transverse valence $u$- and $d$-quark densities.
{\sf Panel A}: $\hat\rho_1^u(|\hat b|)/2$, $\hat\rho_1^d(|\hat b|)$.
{\sf Panel B}: $|\hat b|\hat\rho_1^u(|\hat b|)/2$, $|\hat b|\hat\rho_1^f(|\hat b|)$.
In both panels, the $u$-quark density is halved.  This eliminates the overall $2u$:$1d$ feature of the proton, so that a direct comparison can readily be made between the $|\hat b|$-dependence of the densities.
}
\end{figure}

In order to obtain a picture of proton internal structure, it is instructive to display the valence-quark densities defined by Eq.\,\eqref{fdensity}.  This is done in Fig.\,\ref{Figfdensity}.  The dimensionless quantities we have drawn can be mapped into physical units using:
\begin{equation}
\rho_1^f(|b| =| \hat b|/m_N) = m_N^2 \,  \hat \rho_1^f(|\hat b|)\,,
\end{equation}
\emph{viz}.\ $|\hat b| = 1$ corresponds to $|b| \approx 0.2\,$fm and $\hat\rho_1^f=0.04 \Rightarrow \rho_1^f\approx 1.0/{\rm fm}^2$.

The images in Fig.\,\ref{Figfdensity} provide valuable insights.\vspace*{-1ex}

\begin{enumerate}[label={(\roman*)}]
\item Fig.\,\ref{Figfdensity}\,A reveals that valence $u$-quarks are more likely than valence $d$-quarks to be found near the proton's centre of transverse momentum (CoTM).
    (This is a light-front generalisation of the concept of distance from the centre of mass in nonrelativistic systems.)
    The excess valence $u$-quark density lies on $|\hat b| \leq 1.75 \Rightarrow |b| \lesssim 0.35\,$fm, with $\hat\rho_1^d(0)/\hat\rho_1^u(0) = 0.32$.
    Both the size of the excess and extent of the associated domain are connected with the relative strength of scalar and axialvector diquark correlations within the proton: omitting the axialvector diquark, the $u$-quark excess is larger.

\item Fig.\,\ref{Figfdensity}\,B indicates that, differing from Ref.\,\cite{Cui:2020rmu}, the $u$- and $d$-quark transverse densities have somewhat different mean radii.  This can be quantified by considering
\begin{align}
\label{TransverseRadii}
(\hat \lambda_1^f)^2 & := \frac{\int\! d^2\hat b\, |\hat b|^2 \hat \rho_1^f(|\hat b|)}{\int\! d^2\hat b\, \hat \rho_1^f(|\hat b|)} =   \nonumber \\
& =\left. - 4 \frac{d}{dx} \ln F_1^f(x)\right|_{x=0}\,,
\end{align}
from which we find
\begin{equation}
\hat \lambda_1^u \approx 2.67 \,, \quad
\hat \lambda_1^d = 2.85\,,
\end{equation}
values corresponding to $0.53\,$fm, $0.57\,$fm, respectively.
Using the fit to data in Ref.\,\cite{Kelly:2004hm}, one finds
$\hat \lambda_1^u \approx 3.09$,
$\hat \lambda_1^d = 3.14$.

\item Working from Eq.\,\eqref{FlavourSep}, one readily reconstructs
\begin{equation}
 \rho_1^n = (2/3)(\rho_1^d-\rho_1^u/2)\,;
\end{equation}
hence, considering Fig.\,\ref{Figfdensity}\,A, it is plain that the transverse density associated with the neutron dressed-quark core is negative on $|\hat b|\lesssim 1.75$, thereafter becoming positive and diminishing to zero from above with increasing $|\hat b|$.

\end{enumerate}

\begin{figure}[t]
\includegraphics[clip, width=0.48\textwidth]{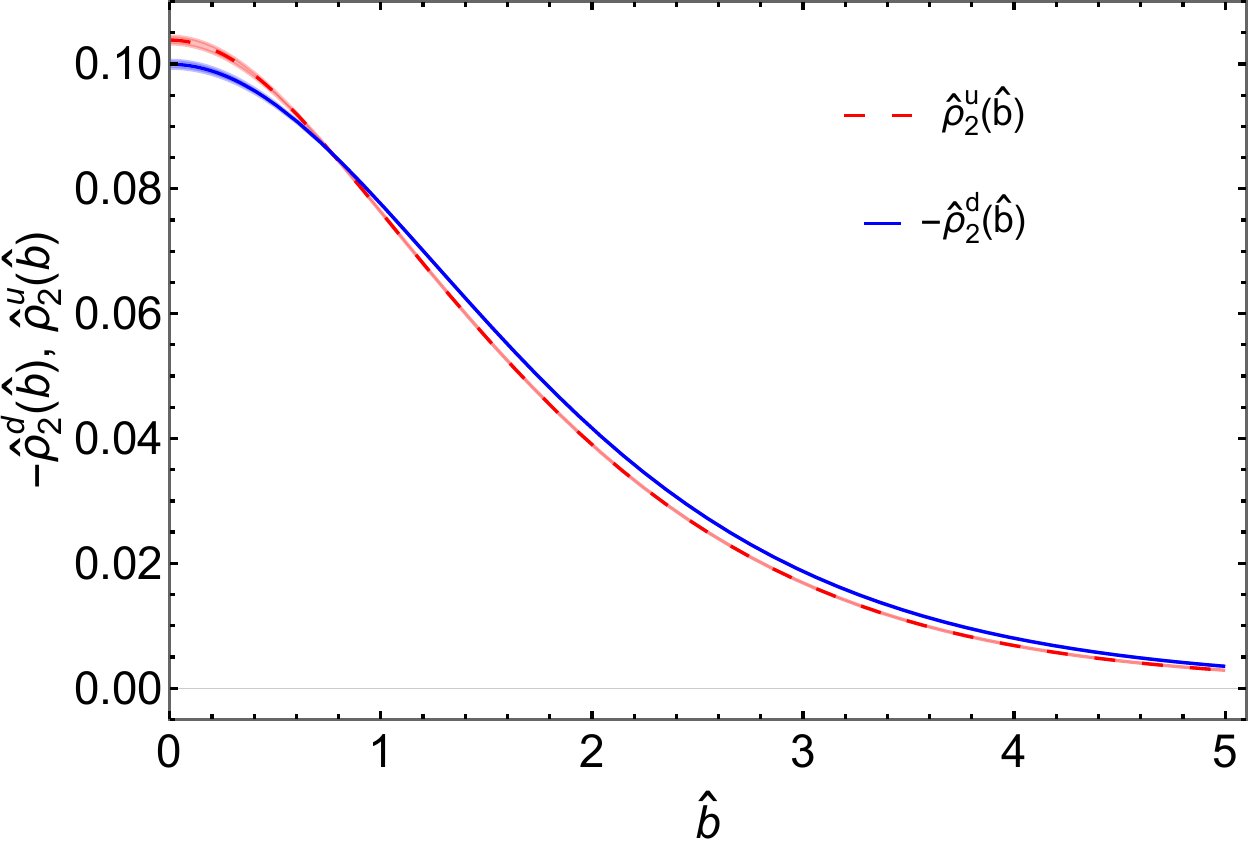}
\vspace*{1ex}

\includegraphics[clip, width=0.48\textwidth]{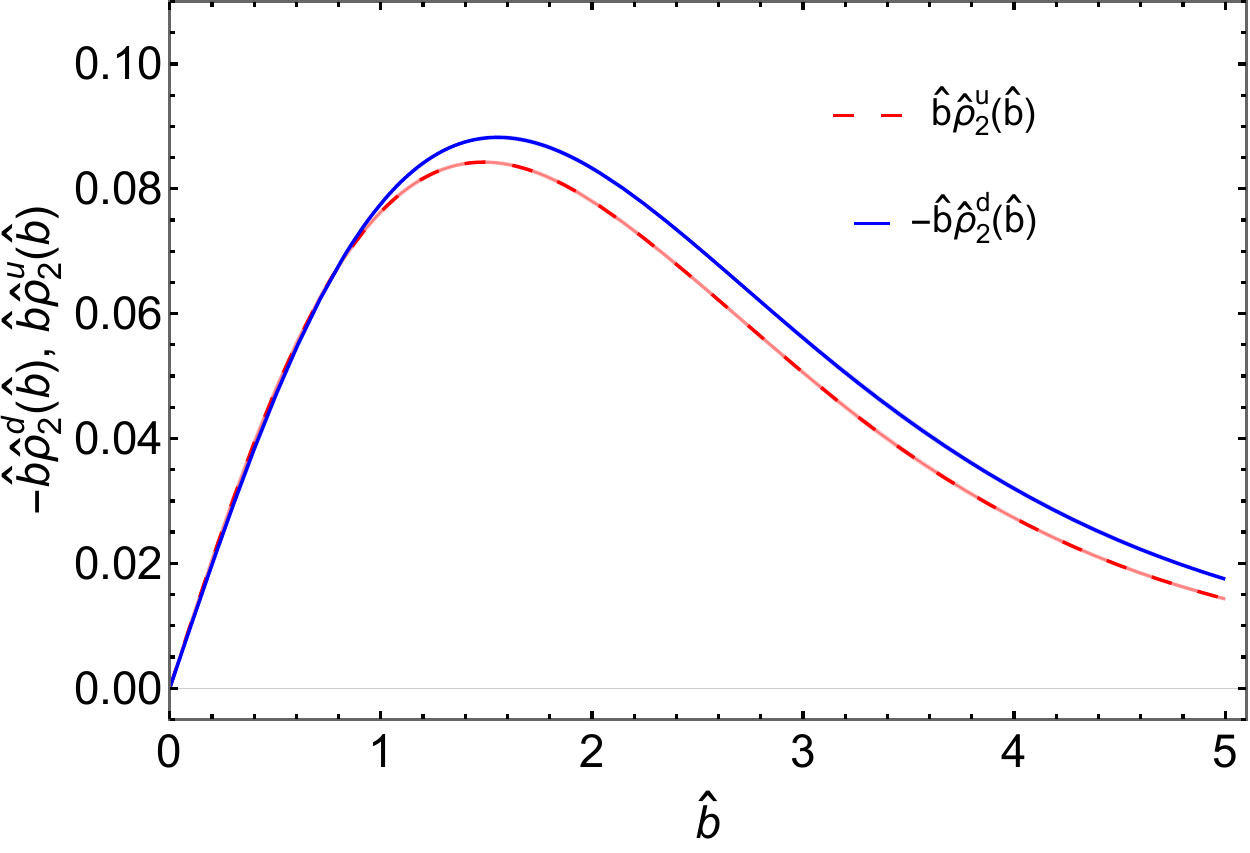}
\caption{\label{FigAMdensity}
Light-front transverse valence $u$- and $d$-quark anomalous magnetisation densities.
{\sf Panel A}: $\hat\rho_1^{u}(|\hat b|)$, $-\hat\rho_1^{d}(|\hat b|)$.
{\sf Panel B}: $|\hat b|\hat\rho_1^{u}(|\hat b|)$, $-|\hat b|\hat\rho_1^{d}(|\hat b|)$.
In these panels, the $u$-quark density is not halved; so, the overall $2u$:$1d$ feature of the proton is preserved and yet the curves differ little.
}
\end{figure}

\subsection{Anomalous magnetisation}
\label{SecAM}
An analogous valence-quark anomalous magnetisation density can also be defined:
\begin{equation}
\label{fdensityF2}
\hat \rho_2^f(|\hat b|) = \int\frac{d^2 \vec{q}_\perp}{(2\pi)^2}\,{\rm e}^{i \vec{q}_\perp \cdot \hat b}
F_2^{f}(Q^2)\,.
\end{equation}
Of course,
\begin{equation}
\kappa_f := F_2^f(0) =  \int d^2\hat b \, \hat \rho_2^f(|\hat b|) \,.
\end{equation}
Our framework yields $\kappa_u=1.74$, $\kappa_d=-1.92$.
Similar values are obtained using the fits to data in Ref.\,\cite{Kelly:2004hm}, \emph{viz}.\ $\kappa_u=1.67$, $\kappa_d=-2.03$.

\begin{enumerate}[label={(\roman*)}]

\item Figure~\ref{FigAMdensity} shows that the valence $d$-quark is magnetically very active within the proton: the $d$-quark profiles are approximately equal in magnitude (opposite in sign) to the $u$-quark profile even though there are two $u$-quarks and only one $d$-quark.
    Within our framework, this is correlated with the fact that, at $\zeta_{\cal H}$, the $u$ valence quark angular momentum in the proton is much less than that of the $d$ valence quark \cite{Cheng:2023kmt, Yu:2024qsd, Yu:2024ovn}.
    Plainly, in a scalar diquark only proton, the $d$ quark would be locked within a magnetically inert $[ud]$ correlation.  So the presence of an axialvector diquark in the proton is crucial to understanding $F_2^d$.

\item Radii may also be associated with the anomalous magnetic moment distributions, defined by analogy with Eq.\,\eqref{TransverseRadii}.  We find
    \begin{equation}
    \hat \lambda_2^u = 3.15 \,,\; \hat \lambda_2^d =3.41\,.
    \end{equation}
    The data parametrisations in Ref.\,\cite{Kelly:2004hm} yield $\hat \lambda_2^u = 3.27$, $\hat \lambda_2^d =3.64$.
    Hence, as plain from Fig.\,\ref{FigAMdensity}\,B, the anomalous magnetisation density connected with the sole valence $d$-quark spreads further from the proton's CoTM than that associated with any given valence $u$-quark.
    This is consistent with $d$ quark contributions being connected with the heavy axialvector diquark piece of the proton wave function.

\end{enumerate}

\subsection{Remarks}
The Faddeev equation in Fig.\,\ref{FigFaddeev} describes the nucleon's dressed-quark core; so, all statements made herein describe that part of a nucleon.
Meson cloud effects may contribute on some bounded domain beyond $|\hat b| \gtrsim 3$ ($\approx 0.6\,$fm), principally modifying features of these distributions at separations from the nucleon's CoTM that lie at the extremity of the range we have typically illustrated and beyond.

The nucleon ground state is just one element in a rich spectrum of baryons.
Consequently, even a complete explanation of nucleon properties reveals only a small piece of a larger picture.
A broader perspective is achieved by also studying nucleon-to-resonance transitions \cite{Carman:2023zke, Achenbach:2025kfx}.
In this context, too, transverse densities may deliver useful insights into EHM and related phenomena \cite{Carlson:2007xd, Tiator:2008kd, Roberts:2018hpf}.

It is worth reiterating that since the elastic and transition form factors are Poincar\'e-invariant and empirically observable, then all associated remarks and conclusions are independent of resolving scale and an observer's frame of reference.
Regarding the light-front-transverse densities, one is treating a projection of Poin\-car\'e-invariant functions; hence, they are frame specific and, in general, resolving scale dependent.

\section{Summary and Outlook}
\label{epilogue}
Beginning with an established Poincar\'e-covariant quark + diquark, $q(qq)$, Faddeev equation approach to nucleon structure [Sect.\,\ref{SecFaddeev}], we developed a refined symmetry preserving current for electron + nucleon elastic scattering [Sect.\,\ref{SecCurrent}].  The interaction current has eight parameters, whose values were chosen so that the $q(qq)$ picture reproduces a few selected features of contemporary $3$-body analyses of nucleon elastic electromagnetic form factors [Sect.\,\ref{Method}].
Notably, the $q(qq)$ Faddeev equation is defined by a kernel whose structure is informed by modern calculations of QCD Schwinger functions and thus incorporates key dynamical features that owe to phenomena underlying emergent hadron mass in the Standard Model \cite{Roberts:2021nhw, Ding:2022ows, Binosi:2022djx, Ferreira:2023fva, Raya:2024ejx}.

Despite being constrained by only a selected subset of $3$-body results, our $q(qq)$ picture reproduces almost all the $3$-body predictions and often form factors in better agreement with available data [Sect.\,\ref{SecResults}].
Of special interest are
the prediction of a zero in $G_E^p/G_M^p$ at $x=Q^2/m_N^2 \approx 11$;
the absence of such a zero in $G_E^n/G_M^n$;
and the prediction of a zero at $x\approx 5.8$ in the proton's $d$-quark Dirac form factor.
Since the current developed herein corrects a collection of errors uncovered in earlier \emph{Ans\"atze}, then, where conflicts exist, our predictions should be viewed as superseding existing studies.

Also of interest are the derived $q(qq)$ results for proton flavour-separated light-front-transverse number and anomalous magnetisation densities as a function of $|\hat b|$, the transverse distance from the proton's centre of transverse momentum.
Notable, for instance, are the predictions that there is an excess of valence $u$-quarks on $|\hat b|\simeq 0$ and the valence $d$-quark is far more active magnetically than either of the valence $u$-quarks.
Within our framework, these features are manifestations of both the presence and relative importance of scalar and axialvector diquark correlations within the proton, which themselves are dynamical consequences of EHM.

With the $q(qq)$ framework reset, it is now appropriate to reanalyse the array of nucleon to resonance transition form factors that have already been studied \cite{Burkert:2017djo, Chen:2018nsg, Lu:2019bjs} and proceed to deliver predictions for those which have not yet been calculated using the QCD-kindred framework \cite{Raya:2021pyr, Liu:2022ndb, Liu:2022nku, Achenbach:2025kfx}.
The $3$-body framework should also be further developed so that it, too, can deliver predictions for such an expanded array of observables.
This would enable additional independent comparisons between the two approaches and opportunities for greater synergy.
One may then move closer to answering a key question; namely, is the quark + fully-interacting diquark picture of baryon structure merely an efficacious phenomenology or does it come close to representing a veracious description of baryons?

\begin{CJK*}{UTF8}{gbsn}
\begin{acknowledgements}
We are grateful to Z.-F.\ Cui (崔著钫) for discussions.
%
Work supported by:
National Natural Science Foundation of China (grant nos.\ 12135007, 12205149);
%
%
and
Natural Science Foundation of Anhui Pro\-vince (grant no.\ 2408085QA028).
\end{acknowledgements}
\end{CJK*}

\begin{small}

\noindent\textbf{Data Availability Statement} Data will be made available on reasonable request.  [Authors' comment: All information necessary to reproduce the results described herein is contained in the material presented above.]
\medskip

\noindent\textbf{Code Availability Statement} Code/software will be made available
on reasonable request. [Authors' comment: No additional remarks.]

\end{small}

\bibliographystyle{elsarticle-num-names}
\bibliography{../../../../CollectedBiB}

\end{document}